\documentclass[twocolumn,showpacs,preprintnumbers,
prb,superscriptaddress]{revtex4-1}%
\usepackage{amssymb}
\usepackage{mathtools}
\usepackage{amsmath}
\usepackage{graphicx}
\usepackage{float}
\usepackage{dcolumn}
\usepackage{bm}
\usepackage{amsfonts}%
\usepackage[usenames, dvipsnames]{color}
\providecommand{\U}[1]{\protect\rule{.1in}{.1in}}
\definecolor{greenish}{rgb}{0.20, 0.65, 0.25}

\newcommand*{\citen}[1]{%
  \begingroup
    \romannumeral-`\x % remove space at the beginning of \setcitestyle
    \setcitestyle{numbers}%
    \cite{#1}%
  \endgroup   
}

\begin{document}
\title{Position Representation of Effective Electron-Electron Interactions in Solids}
\author{T. J. Sj\"{o}strand}
\affiliation{Department of Physics, Division of Mathematical Physics, Lund University, Professorsgatan 1, 22363 Lund, Sweden}
\author{F. Nilsson}
\affiliation{Department of Physics, Division of Mathematical Physics, Lund University, Professorsgatan 1, 22363 Lund, Sweden}
\author{C. Friedrich}
\affiliation{Peter Gr\"{u}nberg Institut and Institute for Advanced Simulation, Forschungszentrum J\"{u}lich and JARA, 52425 J\"{u}lich, Germany}
\author{F. Aryasetiawan}
\affiliation{Department of Physics, Division of Mathematical Physics, Lund University, Professorsgatan 1, 22363 Lund, Sweden}

\date{\today }
\begin{abstract}
An essential ingredient in many model Hamiltonians, such as the Hubbard model, is the effective electron-electron interaction $U$, which enters as matrix elements in some localized basis. These matrix elements provide the necessary information in the model, but the localized basis is incomplete for describing $U$. We present a systematic scheme for computing the manifestly basis-independent dynamical interaction in position representation, $U({\bf r},{\bf r}';\omega)$, and its Fourier transform to time domain, $U({\bf r},{\bf r}';\tau)$. These functions can serve as an unbiased tool for the construction of model Hamiltonians. For illustration we apply the scheme within the constrained random-phase approximation to the cuprate parent compounds La$_2$CuO$_4$ and HgBa$_2$CuO$_4$ within the commonly used 1- and 3-band models, and to non-superconducting SrVO$_{3}$ within the $t_{2g}$ model. Our method is used to investigate the shape and strength of screening channels in the compounds. We show that the O 2$p_{x,y}-$Cu 3$d_{x^2-y^2}$ screening gives rise to regions with strong attractive static interaction in the minimal (1-band) model in both cuprates. On the other hand, in the minimal ($t_{2g}$) model of SrVO$_3$ only regions with a minute attractive interaction are found. The temporal interaction exhibits generic damped oscillations in all compounds, and its time-integral is shown to be the potential caused by inserting a frozen point charge at $\tau=0$. When studying the latter within the three-band model for the cuprates, short time intervals are found to produce a negative potential. 
\end{abstract}

\maketitle

\section{Introduction}
One of the most important quantities in many-electron physics is the screened Coulomb interaction between two electrons, $W$, which is a central quantity entering the Hedin equations.\cite{hedin65} Its asymptotic value ($\omega \to \infty$) equals the bare Coulomb interaction $v$, whereas its static value ($\omega \to 0$) is very much reduced compared to $v$ due to the dynamic screening of the system, embodied by the retarded response. For finite $\omega$, it becomes a complex quantity whose imaginary part can be directly related to the experimentally measured energy-loss spectra.\cite{hedin67} Many quantities and equations are intimately tied to $W$ since the electron self-energy $\Sigma$ is a functional of it. One example is Eliashberg theory of superconductivity,\cite{Eli,Eli2} which for years has been investigated in terms of effective interactions,\cite{ashcroft} and which recently was made parameter free by making use of $W$,\cite{jap} just as in superconducting density functional theory.\cite{scdft0,scdft} A quantity closely related to $W$ is the effective low-energy interaction or partially screened interaction $U$, which excludes screening from a low-energy subspace corresponding to a model Hamiltonian and may be regarded as a dynamical and non-local generalization of the Hubbard on-site repulsion.\cite{downfold,downfold2,Upaper}

In the position representation, $W$ and $U$ are functions of two position variables and time (or frequency): $W(\bf{r,r'};\tau)$, $U(\bf{r,r'};\tau')$, but little is known about the actual shape of these functions. The focus is typically on their matrix elements in some set of orbitals, either because these are needed when calculating other quantities or because they are central objects in Hubbard-like models. However, matrix elements are basis-dependent and, since being projected quantities, do not contain complete information about the screened interaction. We therefore present a systematic scheme which allows for the computation of the position representations of the frequency-dependent $W$ and $U$, manifestly independent of any basis. This provides an unbiased tool to pin down how a suitable model can be constructed in a given periodic solid. A subsequent Fourier transform reveals the full spatiotemporal interactions $W({\bf r},{\bf r}';\tau)$, $U({\bf r},{\bf r}';\tau)$. A space-time point of view may furnish useful complementary insights into the physics problem at hand, like that of high-$T_C$ superconductivity. To illustrate the use of the developed scheme, we compute the screened interactions in the well-known high-temperature superconductor parent compounds La$_2$CuO$_4$ (LCO) and HgBa$_2$CuO$_4$ (HBCO), and for comparison in non-superconducting SrVO$_{3}$, a prototype of correlated metals. 

Shortly after the ground-breaking discovery of high-temperature superconductivity in the doped cuprates\cite{1986} it was realized that standard Bardeen-Cooper Schrieffer (BCS) theory\cite{bcstheory}, based on electron-phonon interaction, could neither account for their elevated critical temperatures nor their anomalous and doping-dependent isotope effect.\cite{notbcs} In the well-underdoped non-superconducting regime, the cuprates share an antiferromagnetic Mott insulating order caused by strong repulsion in the partially filled Cu 3$d$ band,\cite{mott} and the superconducting phase emerges, as a consequence of doping, in the vicinity of a Mott transition. It was, for this reason, early pointed out that the pairing mechanism ought to be mainly of electronic or magnetic origin,\cite{andmott} a viewpoint which is reinforced by the $d_{x^2-y^2}$ symmetry of the superconducting gap.\cite{scalapino,dsym} Unfortunately, despite the progress in the field of strong correlations, there is to this day still no consensus on what mechanism or, rather, interplay of mechanisms best describes this pairing.

The strong correlations of these materials explain the qualitative failure of the local density approximation (LDA), which predicts a metal for the undoped parent compounds. The deceptively simple low-energy electronic structure can be traced back to the CuO$_{2}$ sheet, in which the Cu 3$d_{x^{2}-y^{2}}$ and O 2$p_{x/y}$ orbitals hybridize to form a bonding and an antibonding state.\cite{halffilled} The antibonding state, which has a strong Cu 3$d_{x^{2}-y^{2}}$ weight, forms the half-filled and well-isolated narrow band across the Fermi level in LDA. Indeed, this antibonding band is commonly used to model the low-energy electrons participating in superconductivity and frequently constitutes one of the orbitals in model Hamiltonians.\cite{emery} The additional low-lying oxygen $p$ bands provide a strong screening channel that causes a substantial reduction in the effective interaction.\cite{laurentium}

Many pairing mechanisms have been put forward over the last three decades. Anderson\cite{anderson} suggested that strong short-range repulsive interactions lead to spin-charge separation and that the immense antiferromagnetic superexchange opens up a $d-$wave spin gap, which by kinetic frustration converts to a superconducting gap. The charge fluctuation mechanism dates back to Kohn and Luttinger,\cite{KL} who realized that Friedel oscillations lead to anisotropic pairing in an isotropic electron gas with short-range interactions at low temperatures. Numerical studies within the random-phase approximation (RPA) by Rietschel and Sham later confirmed this for a certain range of electron densities by solving the Eliashberg equation.\cite{RSham} Since spin fluctuations are believed to completely overshadow charge fluctuations at short distances, the latter has not been extensively investigated for the cuprates. It is conceivable that the electron gas results persist in realistic materials, but that the relevant length scale is significantly reduced. Indeed, Kohn and Luttinger argued that a non-spherical Fermi surface can drastically increase $T_C$.\cite{KL} The screened interaction in position representation may furnish a physical insight into this mechanism, not easily accessible from matrix elements alone. 

For the undoped cuprates we consider the famous one- and three-band models and calculate the effective interactions $U_1$ and $U_3$ in the respective low-energy subspace. The metallic band with dominating Cu 3$d_{x^2-y^2}$ weight constitutes the one-band subspace, whereas the three-band subspace also includes two bonding and non-bonding bands of mainly O 2$p_{x,y}$ character.\cite{emery} $U$ does not include the screening of the electrons of the subspace, hence also the screening from the pathological metallic band is excluded, which partly justifies the use of LDA as a starting point. It is worth noting that the charge gap in LCO, which is absent at the LDA level, is opened up within LDA+DMFT when a dynamic $U$ computed using constrained RPA (cRPA) is used, whereas when the static value is used the material remains metallic.\cite{laurentium} The measured gap of 2 eV is almost perfectly reproduced in the three-band model and partly so in the one-band model,\cite{laurentium} which shows that $U$, when calculated within cRPA, indeed embodies dynamical correlation effects required when modeling the undoped cuprates. We also calculate the fully screened interaction $W$ although its interpretation demands some caution. With some justification, it may be thought of as a crude estimation of the screened interaction of the metallic doped system, which could be systematically improved, for instance, by imposing rigid shifts in the LDA band filling.\cite{fermi}

This paper is organized as follows: In Section \ref{Sec:theory} we summarize the theory of the partially and fully screened Coulomb interaction, $U$ and $W$, as well as the RPA and constrained RPA approximations. In Section \ref{Sec:spacetime} the space-time computation of $W({\bf r},{\bf r}';\tau)$ and $U({\bf r},{\bf r}';\tau)$ is described, and their interpretations are emphasized. In Section \ref{Sec:Results} the results for SrVO$_3$, LCO and HBCO are presented and discussed and in Section \ref{Sec:Conclusions} the main findings are summarized. 

\section{Screened Interaction}
\label{Sec:theory}
\label{DFT} 
%SCDFT\cite{scdft,kspert} equations for the superconducting gap $\Delta$.
\subsection{$W$ and RPA}
\label{test}

Before describing the position-space computation of $W(\mathbf{r,r}^{\prime};t-t^{\prime})$ or $W(\mathbf{r,r}%
^{\prime};\omega)$ we recapitulate the definition of $W$ from linear response theory. When applying an arbitrary external perturbation $V_\text{ext}(\mathbf{r},t)$ the induced density is to linear order given by
\begin{align}
\delta \rho (\mathbf{r},t)=\int d{\bf r}^{\prime}dt^{\prime}\chi(\mathbf{r},
t;\mathbf{r}^{\prime},t^{\prime})V_\text{ext}(\mathbf{r}^{\prime},t^{\prime}),
\end{align}
where $\chi$ is the linear density response function. This causes a change in the Hartree potential
\begin{align}
\delta V_H (\mathbf{r},t)=\int d{\bf r}^{\prime}v(\mathbf{r-r}^{\prime})\delta \rho
(\mathbf{r}^{\prime},t),
\end{align}

\noindent which screens the applied perturbation $V_\text{ext}$. The resulting change in the total
potential $\delta V=V_\text{ext} + \delta V_H$ is given by
\begin{align}
\delta V (\mathbf{r},t) &=V_\text{ext}(\mathbf{r},t) \label{arbPert} \\ 
&+\int d{\bf r}_{1} d{\bf r}_2 dt_2 v(\mathbf{r-r}_{1}) \chi(\mathbf{r}_{1},t;\mathbf{r}_{2},t_{2})V_\text{ext}(\mathbf{r}_{2}%
,t_{2}). \nonumber
\end{align}

\noindent Schematically we may write
\begin{align}
\delta V=(1+v\chi)V_\text{ext} = \epsilon^{-1}V_\text{ext} 
\end{align}

\noindent where we recognize that $1+v\chi$ is the inverse dielectric matrix $\epsilon^{-1}$. If we replace our external perturbation with the Coulomb interaction $v(\mathbf{r-r}^{\prime})\delta(t-t^{\prime})=\delta
(t-t^{\prime})/|\mathbf{r}-\mathbf{r}^{\prime}|$, with $(\mathbf{r}^{\prime},t^{\prime})$ treated as a parameter, we arrive at
\begin{align}
W(\mathbf{r,r}^{\prime};\tau)  &\equiv  v(\mathbf{r-r}^{\prime})\delta(\tau)\label{Wrrtime} \\
&+\int d{\bf r}_{1}d{\bf r}_{2}v(\mathbf{r-r}_{1}) \chi(\mathbf{r}_{1},\mathbf{r}_{2};\tau)v(\mathbf{r}_{2}-\mathbf{r}^{\prime}). \nonumber
\end{align}
\noindent This is the definition of the screened interaction in the Hedin equations.\cite{hedin65} The second term, $v\chi v$, which is the screening contribution to $W$, is usually denoted by $W^\mathrm{c}$, a notation we will adopt in the following. We have made use of the fact that $\chi$ depends only on relative time $\tau=t-t'$ for a system with time-independent Hamiltonian. $W({\bf r},{\bf r}';\tau)$ is the effective interaction between two electrons at ${\bf r},t$ and ${\bf r}',t'$ and contains a retarded contribution, $W^\mathrm{c}$, due to the dynamic response of all electrons in the system. Within RPA, this retarded response originates from successive particle-hole excitations caused by the instantaneous interaction between the electrons in the system. The Fourier component of the screened interaction is then calculated from the following equation:
\begin{align}
W(\mathbf{r,r}^{\prime};\omega) &=v(\mathbf{r-r}^{\prime})\label{Wspace} \\ &+\int d{\bf r}_{1}%
d{\bf r}_{2}v(\mathbf{r-r}_{1}) \nonumber  \chi(\mathbf{r}_{1},\mathbf{r}_{2};\omega)v(\mathbf{r}_{2}-\mathbf{r}^{\prime}). 
\end{align}

\noindent The screened Coulomb interaction, $W$, is uniquely determined by the linear density response function $\chi=\delta\rho/\delta\varphi$. We can introduce the irreducible polarization propagator $P$, which may be thought of as the linear density response function with respect to the total field, $P=\delta \rho/\delta V$. It then follows from the chain rule that
\begin{equation}
\chi=P+Pv\chi
\end{equation}

\noindent and 
\begin{equation}
W=v+v\chi v=v+vPW = v+W^\mathrm{c}.
\end{equation}

\noindent In the random-phase approximation (RPA) the polarization propagator is approximated by the response function of a noninteracting system $\chi^0$,\cite{hedin65} so that the response function takes the form
\begin{align}
\chi^\text{RPA}=\chi^0 + \chi^0 v \chi^\text{RPA}\label{RPAeq},
\end{align}

\noindent where 
\begin{align}
&  \chi^{0}(\mathbf{r},\mathbf{r}^{\prime};\omega)=2\sum_{{\bf k}n}^\text{occ} \sum_{{\bf k}'n'}^\text{unocc}\chi_{n\mathbf{k},n^{\prime}\mathbf{k}^{\prime}}^{0}(\mathbf{r}%
,\mathbf{r}^{\prime}\!;\omega)\label{pol}\\
&  \chi_{n\mathbf{k},n^{\prime}\mathbf{k}^{\prime}}^{0}(\mathbf{r},\mathbf{r}%
^{\prime}\!;\omega)=\phi_{n\mathbf{k}}^{\ast}(\mathbf{r})\phi_{n^{\prime
}\mathbf{k}^{\prime}}(\mathbf{r})\phi_{n\mathbf{k}}(\mathbf{r}^{\prime}%
)\phi_{n^{\prime}\mathbf{k}^{\prime}}^{\ast}(\mathbf{r}^{\prime})\nonumber
\end{align}
\[
\times\left(  \frac{1}{\omega\!+\!\varepsilon_{n\mathbf{k}}\!\!-\!\varepsilon
_{n^{\prime}\mathbf{k}^{\prime}}\!\!+\!i0^{+}}-\frac{1}{\omega\!-\!\varepsilon
_{n\mathbf{k}}\!\!+\!\varepsilon_{n^{\prime}\mathbf{k}^{\prime}}\!\!-\!i0^{+}%
}\right)  
\]

\noindent is equivalent to the well-known Lindhard formula.\cite{lindhart} Here, $\phi_{n{\bf k}}$ and $\varepsilon_{n{\bf k}}$ are paramagnetic eigenfunctions and eigenenergies, typically obtained using density functional theory (DFT). ${\bf k}$ is restricted to the first Brillouin zone. The factor of two is due to summing over the two identical spin contributions, and the two sums are restricted to occupied (occ) and unoccupied (unocc) states respectively. Note that Eq. \eqref{pol} describes the time-ordered polarization function, which means that the resulting screened interaction is not the retarded, but the time-ordered $W$. One can recover the retarded $W$ by multiplying the imaginary part of the time-ordered $W$ by a factor of $\text{sign}(\omega)$.
\begin{figure}
\includegraphics[width=.99\linewidth]{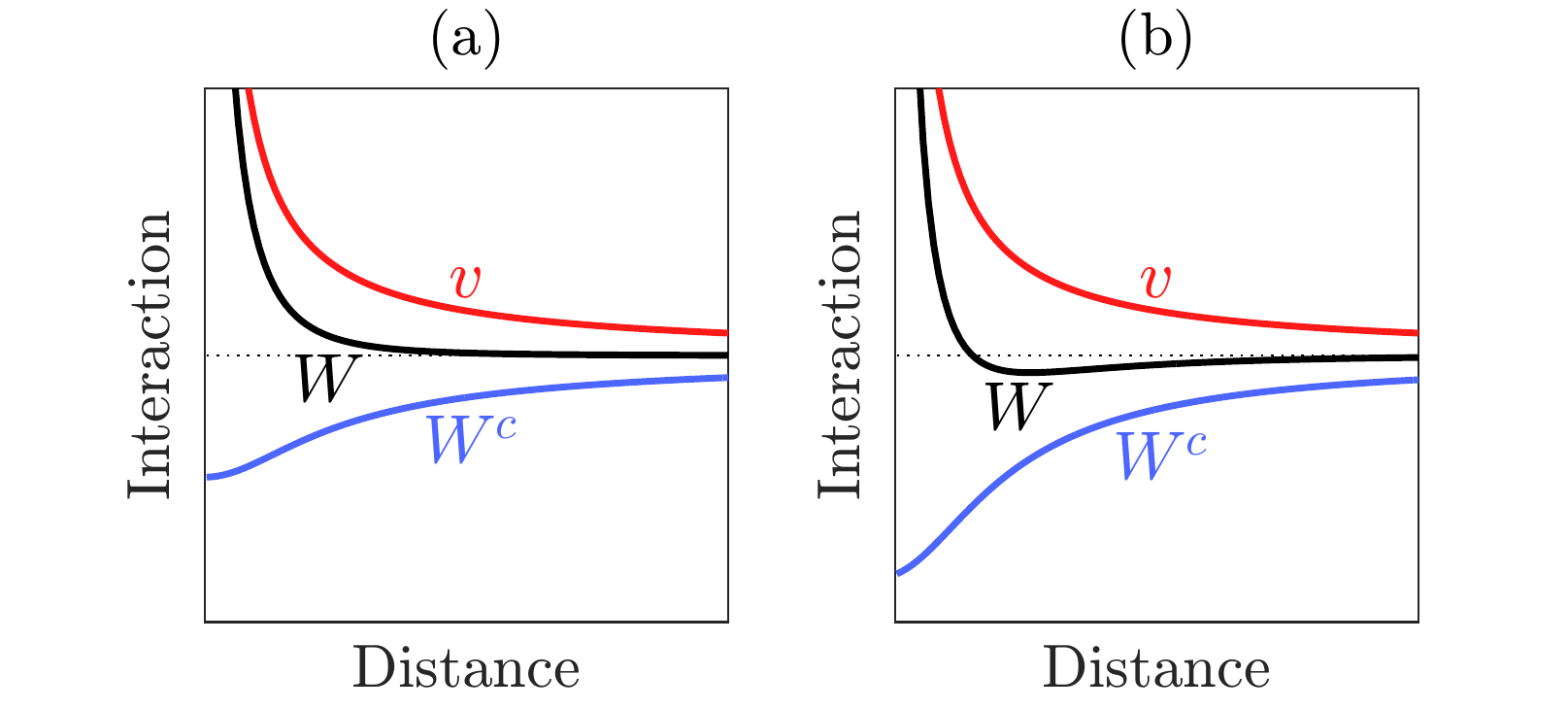}
\caption{Qualitative illustration of the screened interaction $W=v+W^c$ and its constituents $v$ and $W^c$ in the static limit. (a) Shallow screening hole. (b) Deep screening hole.}\label{schematicWplot}
\end{figure}

A qualitative and simplified depiction of $W$ is presented in Fig. \ref{schematicWplot}. By increasing the depth of the screening hole, the effective interaction is reduced and can even turn negative at certain distances. The terms attraction and repulsion, however, have to be used with caution since they originate from situations where the interaction is radially monotonous and thus either attractive or repulsive throughout. Still, we adopt the term attraction if we, for a given $\mathbf{r}'$, identify a negative minimum of the interaction at ${\bf r}$ (local attraction) towards which the classical force field is pointing. 

As can be seen from Fig. \ref{schematicWplot}, at very short distances to ${\bf r}'$, the force field is always pointing outwards, which gives a local repulsion. This can be understood intuitively since for $\mathbf{r}\rightarrow\mathbf{r}'$ there is not sufficient charge in the region between $\mathbf{r}$ and $\mathbf{r}'$ to create screening holes that could compensate or overcompensate the Coulomb repulsion. The screening inherently depends on the electron density in the solid. Different materials will have different screening properties and therefore also different shapes of $W(\mathbf{r},\mathbf{r}')$. The placement of the point charge will therefore also matter. If it is put at the position of a nucleus, especially of an atomic species which is an effective ''screener'', a much more reduced $W$ emerges at short distances than from a point charge in between two nuclei. 
\subsection{$U$ and cRPA}
\label{UandcRPA}
To determine the effective interaction of a low-energy model, we use the cRPA method,\cite{downfold,crpa1} in which the Hilbert space is divided into a low- and a high-energy subspace, $\mathcal{D}$ and $\mathcal{R}$. The polarization function is now decomposed into two terms, $P\!=\!P^\mathrm{d}\!+\!P^\mathrm{r}$. $P^\mathrm{d}$ describes polarization processes within the low-energy subspace $\mathcal{D}$ whereas $P^\mathrm{r}$ accounts for the rest of the polarizations, i.e., those within the $\mathcal{R}$ subspace as well as those between the subspaces. By defining 
\begin{equation}
W^{\text{r}}=v+vP^{\text{r}}W^{\text{r}}
\end{equation}
it can be shown that\cite{downfold}
\begin{equation}
W\!=\!W^{\text{r}}\!+\!W^{\text{r}}P^{\text{d}}W,
\end{equation}
which allows us to interpret $W^{\text{r}}$ as the
effective \textquotedblright bare\textquotedblright\ interaction in
$\mathcal{D}$, a non-local and dynamical generalization of the Hubbard on-site
repulsion.\cite{downfold2} So, 
\begin{align}
U(\mathbf{r},\mathbf{r}^{\prime};\omega)\equiv W^\text{r}({\bf r},{\bf r}';\omega).
\end{align}
As in the case of $W$, we can write
$U\!=\!v\!+\!U^{\text{c}}$, where $U^{\text{c}}\!=\!v\chi^{\text{r}%
}v$ and $\chi^{\text{r}}\!=\!P^{\text{r}}\!+\!P^{\text{r}}v\chi^{\text{r}}$. The low-energy subspace in the Hubbard model usually corresponds to a narrow band with strong correlations, so RPA is not expected to work well. However, when computing $U$ for the model, the polarization channels within the low-energy subspace are removed from Eq. \eqref{pol}, so that it is justifiable to constrain the RPA to compute
$P^\text{r}=\chi^{\text{r}0}$. 

The physics lies in the choice of the low-energy model subspace. For the low-energy bands of the cuprates, which are entangled, we use the ''disentanglement'' schemed developed in Ref. \citen{disen} and define the $\mathcal{D}$ subspace in terms of maximally localized Wannier functions\cite{maxloc} and the $\mathcal{R}$ subspace as the orthogonal space. Computational details for the calculation of $U$ in the cuprates and in SrVO$_3$ are provided in App. \ref{appendixB}. 

\section{Position Representation}
\label{Sec:spacetime}  
This section deals with the computation of $W$ in position representation (Eq. \eqref{Wspace}) and its interpretation in time domain. Any expression for $W$ has an analogue for $U$  obtained by replacing $\chi^{0}$ with $\chi^{\text{r}0}$.

\subsection{Product Basis}
To expand the polarization $\chi^0$ and response function $\chi^\text{RPA}$ we need a set of two-particle basis functions in the form of a product basis $\{B_{\alpha}^{\mathbf{k}}\}$. This basis can be tailored to give a complete representation of $\chi^0$ and be optimized such that a minimal number of basis functions is needed\cite{gunnarsson,gunnarsson2}
\begin{equation}\label{chi0}
\chi^0(\mathbf{r},\mathbf{r}';\omega) = \sum_{\mathbf{k},\alpha \beta} B_{\alpha}^{\mathbf{k}}(\mathbf{r}) \chi^{0\mathbf{k}}_{\alpha \beta}(\omega)  B_{\beta}^{\mathbf{k}*}(\mathbf{r}').
\end{equation}
From $\chi^\text{RPA}=\chi^0+\chi^0v\chi^0+...$, it is clear that the product basis is also complete for representing $\chi^\text{RPA}(\mathbf{r},\mathbf{r}';\omega)$, since $v$ is always sandwiched between two $\chi^0$ so that it is immaterial whether the product basis is complete or not for $v$. In other words, only the projection of $v$ in the subspace of $\chi^0$ is needed. In fact, the product basis constructed for $\chi^0$ is in general far from complete for representing $v(\mathbf{r}-\mathbf{r}')$. Since $W=v+v\chi^\text{RPA} v$ within RPA, this implies that the product basis in general cannot be used for a complete representation of $W(\mathbf{r},\mathbf{r}';\omega)$. The way around this problem is explained in the following. 

\begin{figure}
\includegraphics[width=.98\linewidth]{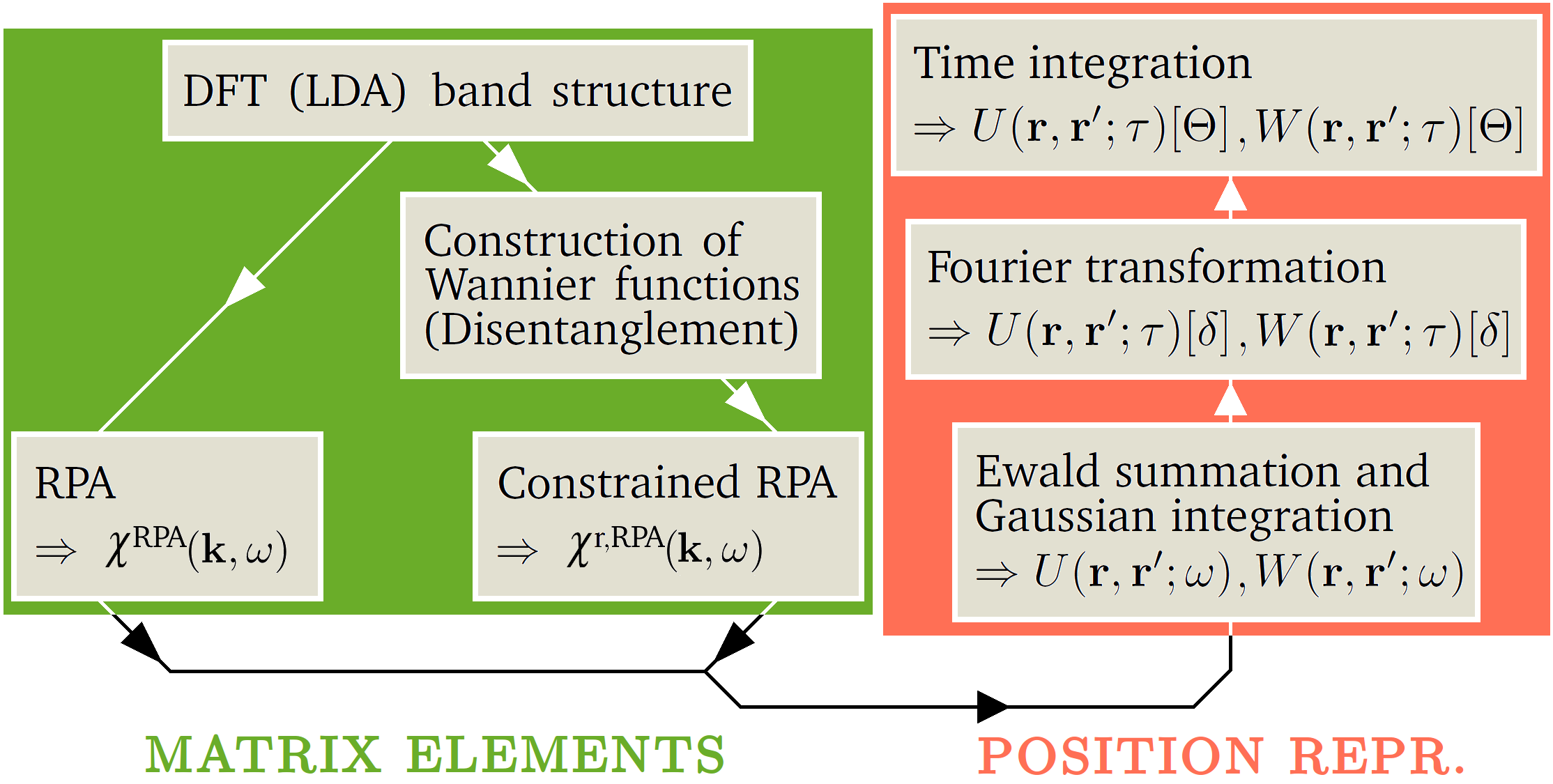}
\caption{Schematics outlining the generation of matrix elements of $\chi$ (green box) used for the computation of $W$ and $U$ in position representation (red box). $W[\delta]$, $U[\delta]$ and $W[\Theta]$, $U[\Theta]$ are defined in section \ref{time}.}\label{schemapic}
\end{figure}

\subsection{$W$ and $U$ in Position Space}\label{WUspace}
Figure \ref{schemapic} shows the steps involved to obtain the matrix elements 
$\chi_{\alpha\beta}^{\text{RPA}}(\mathbf{k};\omega)$ within RPA and $\chi_{\alpha\beta}^{\text{r},\text{RPA}}(\mathbf{k};\omega)$ within cRPA. These matrix elements together with the product basis completely determine $W({\bf r},{\bf r}';\omega)$. 
Since $W$ partly consists of the bare Coulomb interaction $v$, which is known analytically, it is sufficient to find an expression for $W^\mathrm{c}$. Schematically, if we let matrix elements be underlined, Eq. \eqref{chi0} reads $\chi^0 = B {\underline\chi}^0 B^*$. Similarly, within RPA it also holds that $\chi^\text{RPA} = B \underline{\chi}^\text{RPA} B^*$, which implies, together with Eq. \eqref{Wspace}, that $W^\mathrm{c}= v \chi^\text{RPA} v = (v B) \underline{\chi}^\text{RPA} (vB)^*$. We have now obtained a basis which is complete for $W^\mathrm{c}$. Explicitly, 
\begin{align}
W^{\mathrm{c}}(\mathbf{r},\mathbf{r}^{\prime};\omega)  &  =\label{vB} \sum_{ \mathbf{k} ,\alpha\beta}\mathcal{I}_{\alpha}^{\mathbf{k}%
}(\mathbf{r})\chi_{\alpha\beta}^{\text{RPA}}(\mathbf{k};\omega)\mathcal{I}%
_{\beta}^{\mathbf{k}\ast}(\mathbf{r}^{\prime})
\end{align}
where 
\begin{align}
&\mathcal{I}_{\alpha}^{\mathbf{k}}(\mathbf{r})    = \label{Ika} \int d\mathbf{r}%
_{1}v(\mathbf{r}-\mathbf{r}_{1})B_{\alpha}^{\mathbf{k}}(\mathbf{r}_{1}), \\
&\chi_{\alpha\beta}^{\text{RPA}}(\mathbf{k}%
;\omega)=\langle\tilde{B}_{\alpha}^{\mathbf{k}}|\chi^{\text{RPA}}%
(\omega)|\tilde{B}_{\beta}^{\mathbf{k}}\rangle . \label{chiab}
\end{align}
\eqref{vB}-\eqref{chiab} are the main equations for obtaining $W$ in position representation. In general, the set of functions $\{\tilde{B}_\alpha^{\bf k} \}$ is biorthogonal to the set $\{B_\alpha^{\bf k} \}$ and fulfills Eqs. \eqref{b1}-\eqref{b3}.

After having obtained all matrix elements $\chi_{\alpha \beta}^\text{RPA}({\bf k};\omega)$, what remains is to calculate the basis-dependent integrals $\mathcal{I}_\alpha^{\bf k}({\bf r})$ as well as including the $\boldsymbol{\Gamma}$-point contribution in a suitable way. We will explain both steps in the following, but first we present the product basis, constructed in the SPEX code, which has been used in this work. 

%\fredrik{checked until here}
\subsubsection{Mixed Product Basis}\label{mpb}
The mixed product basis is an extension of the optimized product basis within the full-potential linearized augmented plane-wave (FLAPW) method,\cite{takao,FLAPW} where space is separated into spherical ''muffin tin'' (MT) spheres around each atom as well as the ''interstitial region'' (IR), which constitutes the remaining region of space. In the MT spheres the product basis functions $B_{aLMP}^{\mathbf{k}}(\mathbf{r})=b_{aLP}(r)Y_{LM}(\hat{{\bf r}})$ are constructed from products of the MT functions of the LAPW basis. Here, $a$ is an orbital index, $L$ and $M$ denote the orbital and magnetic quantum numbers, respectively, and $P$ is an index for different radial functions. In the IR, products of plane waves, which are themselves plane waves $B_{\mathbf{G}}^{\mathbf{k}}(\mathbf{r}%
)=\text{e}^{i(\mathbf{k}+\mathbf{G})\cdot\mathbf{r}}/\sqrt{\Omega}$ are constructed, where $\Omega$ is the unit cell volume. 
The resulting \textquotedblright mixed product basis\textquotedblright\ functions\cite{FLAPW}
\begin{align}
\{B_{\alpha}^{\mathbf{k}}\} &=\{B_{aLMP}^{\mathbf{k}},B_{\mathbf{G}}%
^{\mathbf{k}}\}, \label{b1} \\
\langle B_{\alpha}^{\mathbf{k}}|\tilde{B}_{\beta}^{\mathbf{k}}
\rangle &=\delta_{\alpha\beta} , \label{b2} \\
\sum_{\alpha}|B_{\alpha}^{\mathbf{k}
}\rangle\langle\tilde{B}_{\alpha}^{\mathbf{k}}| & =\mathbf{1}. \label{b3}
\end{align}
are either non-zero only in the MT spheres or in the IR. Eq. \eqref{b3} holds in the subspace of $\chi^\text{RPA}$.
\subsubsection{Muffin-tin Contribution}
We start our position space reconstruction by considering the MT spheres, where $\alpha=a,L,M,P$. By defining
\begin{align}
{\bf r}_1 = {\bf r}_a + {\bf a},
\end{align}
where ${\bf r}_a$ is confined to a MT of radius $R_a$ and ${\bf a}$ is the vector pointing to the atomic centre of $a$, Eq. \eqref{Ika} can be re-expressed as 
\begin{align}
&\mathcal{I}_{\alpha}^{\mathbf{k}}(\mathbf{r})=\int_{R_a }%
d{\bf r}_{a}\sum_{\mathbf{T}}\frac{\text{e}^{i\mathbf{k}\cdot ({\bf a} +\mathbf{T})}%
}{|\mathbf{r}_a+\mathbf{a}+\mathbf{T}-\mathbf{r}|}B_{\alpha
}^{\mathbf{k}}(\mathbf{r}_a), \\
&B_{\alpha}^{\mathbf{k}}(\mathbf{r}_a)=b_{\alpha}(r_{a})Y_{LM}(\hat{\mathbf{r}}_{a}).
\end{align}
Here we made use of Bloch's theorem and the sum runs over all lattice vectors $\{ \mathbf{T} \}$. However, 
$\mathcal{I}_\alpha^{{\bf k}}({\bf r})$ does not converge for a finite sum over ${\bf T}$ due to the long-range integrand, so we perform Ewald summation to resolve this issue (red box in Fig. \ref{schemapic}). For ${\bf k}\neq {\bf \Gamma}$ and with $\mathbf{q}=\mathbf{k}+\mathbf{G}$, where $\mathbf{k}$ is restricted to the first Brillouin zone and $\mathbf{G}$ is a reciprocal lattice vector, Ewald's formula reads\cite{ewald}
\begin{align}
\sum_{\mathbf{T}}\frac{\text{e}^{i\mathbf{k}\cdot\mathbf{T}}}{|\mathbf{r}%
-\mathbf{r}_{1}-\mathbf{T}|} &  =\frac{4\pi}{\Omega}  \sum
_{\mathbf{G} }\frac{\text{e}^{-q^{2}/4\gamma^{2}}}{q^{2}}%
\text{e}^{i\mathbf{q}\cdot(\mathbf{r}-\mathbf{r}_{1})}    \\
&  +\gamma\sum_{\mathbf{T}}\frac{\text{erfc}(\gamma|\mathbf{r}-\mathbf{r}%
_{1}-\mathbf{T}|)}{\gamma|\mathbf{r}-\mathbf{r}_{1}-\mathbf{T}|}%
\text{e}^{i\mathbf{k}\cdot\mathbf{T}}, \nonumber
\end{align}
 For $\gamma \to 0$, the real-space sum is recovered, and, for $\gamma \to \infty$, the second term vanishes and the real-space sum is replaced by a summation in reciprocal space. For a properly chosen $\gamma$, however, the expression is short-ranged in both $|\mathbf{r}-\mathbf{T}|$ and $q$. 

We separate $\mathcal{I}_{\alpha
}^{\mathbf{k}}(\mathbf{r})$ into $\mathcal{I}_{\alpha
}^{\mathbf{k}(1)}(\mathbf{r})$ and $\mathcal{I}_{\alpha
}^{\mathbf{k}(2)}(\mathbf{r})$, resulting from the sums over $\mathbf{G}$ and $\mathbf{T}$ respectively. We define $A_{\gamma}(q)\equiv (4\pi/\Omega)\text{exp}(-q^{2}/4\gamma^{2})/q^{2}$ and make a plane-wave expansion in spherical harmonics%\cite{sakurai}
\begin{align}
\text{e}^{-i{\bf q}\cdot {\bf r}_a}=  4\pi \sum_{L=0}^\infty (-i)^L j_L(qr_a) \sum_{M=-L}^L Y_{LM}^*(\hat{{\bf r}}_a)Y_{LM}(\hat{{\bf q}}),
\end{align}
where $j_L$ are the spherical Bessel functions. This yields for the first term
\begin{align}
\mathcal{I}_{\alpha}^{\mathbf{k}(1)}(\mathbf{r}) &  =4\pi(-i)^{L}\int_0^{R_a}dr_{a}r_{a}^{2}b_{\alpha}(r_{a
}) \\
&  \times \sum_{\mathbf{G}}A_{\gamma}(q)j_{L}%
(qr_{a})Y_{LM} (\hat{\mathbf{q}})\text{e}^{i\mathbf{q}%
\cdot\mathbf{r}} \text{e}^{-i{\bf G}\cdot {\bf a}} .\nonumber
\end{align} 
Introducing $\mathbf{r}_{a{\bf T}}\!=\!\mathbf{r}%
\!-\!\mathbf{a}\!-\!\mathbf{T}$, the second term, $\mathcal{I}_{\alpha
}^{\mathbf{k}(2)}(\mathbf{r})$, diverges if $|\mathbf{r}%
_{a}-\mathbf{r}_{a{\bf T}}|\rightarrow0$. To resolve this issue we make use of the expansion %\tor{(cite)}
\begin{align}
\frac{\text{erfc}(\gamma|\mathbf{r}_{a}-\mathbf{r}_{a{\bf T}}|)}{\gamma
|\mathbf{r}_{a}-\mathbf{r}_{a{\bf T}}|} &  =\sum_{L=0}^{\infty}\frac{4\pi}%
{2L+1} \left[  \frac{r_{<}^{L}}{\gamma r_{>}^{L+1}} - g_{L}(r_{a
}, r_{a{\bf T}})\right] \nonumber \\
&  \times\sum_{M=-L}^{L} Y_{LM}^{\ast}({\hat{\mathbf{r}}}_{a}%
)Y_{LM}({\hat{\mathbf{r}}}_{a{\bf T}}),
\end{align}
where $r_{<}=\min (r_{a},\!r_{a{\bf T}})$ and $r_{>}=\max (r_{a},\!r_{a{\bf T}})$. Note that the majority of the terms, corresponding to translations ${\bf T}$ that cause no divergence, can be integrated without the use of this expansion. For brevity, we here keep the expansion in all terms, and arrive at
\begin{align}
 & \mathcal{I}_{\alpha}^{\mathbf{k}(2)}(\mathbf{r})   = \frac{4\pi\gamma
}{2L + 1} \int_{0}^{R_a} dr_{a}r_{a}^{2}b_{\alpha
}(r_{a})\\
&  \times  \sum_{\mathbf{T}} \left[  \frac{r_{<}^{L}}{\gamma r_{>}^{L+1}}-g_{L}(r_{a},r_{a{\bf T}}%
) \right]  Y_{LM}({\hat{\mathbf{r}}}_{a{\bf T}})\text{e}%
^{i\mathbf{k}\cdot(\mathbf{a}+\mathbf{T})}. \nonumber
\end{align}
The coefficients $g_L$ are computed as
\begin{align}
&  \frac{4\pi}{2L+1}g_{L} (r_{a},r_{a{\bf T}})Y_{LM} ({\hat{\mathbf{r}}}_{a{\bf T}})\\
&  =\int d\Omega_{a} \frac{\text{erf}(\gamma|\mathbf{r}_{a
} -\mathbf{r}_{a{\bf T}}|)}{\gamma|\mathbf{r}_{a} -\mathbf{r}_{a{\bf T}}%
|}Y_{LM} ({\hat{\mathbf{r}}}_{a}) \nonumber
\end{align}
using Gaussian integration, meaning that any angular integral $\int d\Omega f(\Omega)$ is replaced by $\sum_i w_i f(\Omega_i)$ where the weights $w_i$ are tabulated and independent of $f$. In particular, we used 114 cubic directions $\Omega_i$, which yields exact results for angular momentum components $L\leq 15$.\cite{ferdibook}
\subsubsection{Interstitial Contribution}
We now consider the IR, where $\alpha={\bf G}$. By extending $B_{\mathbf{G}}^{\mathbf{k}}(\mathbf{r})=
\text{e}^{i\mathbf{q}\cdot\mathbf{r}}/\sqrt{\Omega}
$ to all of space and subtracting the muffin-tin contribution, we can write
\begin{align}
\mathcal{I}_{\mathbf{G}}^{\mathbf{k}}(\mathbf{r}) &  =\int\!d{\bf r}_{1}\frac
{1}{|\mathbf{r}_{1}\!-\!\mathbf{r}|}B_{\mathbf{G}}^{\mathbf{k}}(\mathbf{r}%
_{1})\\
&  -\sum_{\mathbf{a}}\!\int_{R_a}\!d{\bf r}_{a
}\!\sum_{\mathbf{T}}\!\frac{\text{e}^{i\mathbf{k}\cdot({\bf a}+\mathbf{T})} \text{e}^{i {\bf G} \cdot {\bf a}} }%
{|\mathbf{r}_{a}\!+\!\mathbf{a}\!+\!\mathbf{T}\!-\!\mathbf{r}%
|}B_{\mathbf{G}}^{\mathbf{k}}(\mathbf{r}_{a}), \nonumber
\end{align}
where we have made use of the fact that $B_{\mathbf{G}}^{\mathbf{k}}(\mathbf{r}_{a}+\mathbf{T})=\text{e}^{i\mathbf{k}\cdot\mathbf{T}} B_{\mathbf{G}}^{\mathbf{k}}(\mathbf{r}_{a})$.
The first term reads 
\begin{align}
\mathcal{I}_{\mathbf{G}}^{\mathbf{k}(0)}(\mathbf{r})=\frac{4\pi}{\sqrt{\Omega
}q^{2}}\text{e}^{i\mathbf{q}\cdot\mathbf{r}}.
\end{align}
We divide the rest into
$\mathcal{I}_{\mathbf{G}}^{\mathbf{k}(1)}(\mathbf{r})+\mathcal{I}_{\mathbf{G}%
}^{\mathbf{k}(2)}(\mathbf{r})$ from both terms in the Ewald summation in the same manner as before, and analogously we obtain
\begin{align}
\mathcal{I}_{\mathbf{G}}^{\mathbf{k}(1)}(\mathbf{r})\! &  =\!-4\pi
\!\sum_{\mathbf{a}}\!\int_{0}^{R_a}\!\!\!dr_{a
}r_{a}^{2} \sum_{\mathbf{G}^{\prime}}\!A_{\gamma}(q')j_{0}(|\mathbf{G\!-\!G}^{\prime}|r_{a}) \nonumber \\
&  \times\frac{1}{\sqrt{\Omega}}\text{e}^{i\mathbf{q}^{\prime}\cdot\mathbf{r}%
}\text{e}^{i(\mathbf{G}-\mathbf{G}^{\prime})\cdot\mathbf{a}},\\
\mathcal{I}_{\mathbf{G}}^{\mathbf{k}(2)}(\mathbf{r})\! &  = -\frac{(4\pi
)^{2}\gamma}{\sqrt{\Omega}}\!\sum_{\mathbf{a}}\!\int_{0}^{R_a}\!dr_{a}r_{a}^{2}\!\sum_{\mathbf{T}}\sum_{L=0}^{\infty}\!\frac{i^{L}}{2L\!+\!1} \nonumber \\
&  \times\left[  \frac{r_{<}^{L}}{\gamma r_{>}^{L+1}}\!-\!g_{L}(r_{a
},r_{a{\bf T}})\right]  \!j_{L}(qr_{a}) \nonumber \\
&  \times\sum_{M=-L}^{L}Y_{LM}({\hat{\mathbf{r}}}_{a{\bf T}})Y_{LM}^{\ast}%
(\hat{\mathbf{q}})\text{e}^{i\mathbf{q}\cdot(\mathbf{a}+\mathbf{T}%
)},
\end{align}
where ${\bf q}'={\bf k}+{\bf G}'$. Terms in $\mathcal{I}_{\mathbf{G}}^{\mathbf{k}(1)}$ with $L>4$ are very small and excluded in this work. 

\subsubsection{${\bf \Gamma}$-Point Contribution}
What is left at this point is to calculate the ${\boldsymbol \Gamma}$-point contribution to Eq. \eqref{vB}, which requires special treatment since the bare interaction $v$ diverges as $1/k^2$ for $k\!\to\! 0$. In SPEX, the divergence is treated analytically by rotating to the Coulomb eigenbasis\cite{CoulBas}
\begin{align}
E_{\mu}^{{\bf k}}(\mathbf{r})\!=\!\sum_{\alpha}T_{\mu
\alpha}^{{\bf k}}B_{\alpha}^{{\bf k}}(\mathbf{r}).
\end{align}
When $k\to 0$, $E_{\mu=1}^{{\bf k}}(\mathbf{r}) \to 1/\sqrt{\Omega}$ corresponds to the
divergent eigenvalue of $v$ and the matrix element $W_{\mu=1,\nu=1}^{\mathrm{c}}%
({\bf k};\omega)$, which diverges like $1/k^{2}$, just shifts $W^{\mathrm{c}}(\mathbf{r},\mathbf{r}^{\prime
};\omega)$ uniformly to leading order.\cite{FLAPW}
$W_{\mu=1,\nu>1}%
^{\mathrm{c}}({\bf k};\omega)$ and $W_{\mu>1,\nu=1}^{\mathrm{c}}({\bf k};\omega)$ diverge only like
$1/k$ and are much smaller and, for this reason, neglected in this work. This simplification corresponds to making $W^{\mathrm{c}}$ block diagonal in the Coulomb basis. The large block $W_{\mu>1,\nu>1}^{\mathrm{c}}({\bf k};\omega)$ does not contain any divergence, and we therefore rotate it back to the mixed product basis. We then get the ${\bf \Gamma}$-point contribution to $W^{\mathrm{c}}$:
\begin{align}
W_{{\bf k}={\bf 0}}^{\mathrm{c}}(\mathbf{r},\mathbf{r}^{\prime};\omega)  
&=\int_{{\bf \Gamma}}d\mathbf{k}E_{1}^{\mathbf{k}}(\mathbf{r}%
)W_{11}^{\mathrm{c}}(\mathbf{k};\omega)E_{1}^{\mathbf{k}\ast}%
(\mathbf{r}^{\prime}) \nonumber \\
&+\sum_{\alpha\beta}\tilde{\mathcal{I}}_{\alpha}^{{\bf 0}
}(\mathbf{r})\chi_{\alpha\beta}^{\text{RPA}}({\bf 0};\omega)\tilde{\mathcal{I}}%
_{\beta}^{{\bf 0} *}(\mathbf{r}^{\prime}) \label{gamma} ,
\end{align}
where
\begin{align}
\tilde{\mathcal{I}}_{\alpha}^{{\bf 0}}(\mathbf{r})=\sum_{\mu>1}\left(
T^{-1}\right)_{\alpha \mu}^{{\bf 0}}\int d\mathbf{r}_{1}v(|\mathbf{r}%
-\mathbf{r}_{1}|)E_{\mu}^{{\bf 0}}(\mathbf{r}_{1}).
\end{align}
$\tilde{\mathcal{I}}_\alpha^{\bf 0}$ is calculated in the same way as $\mathcal{I}_\alpha^{\bf k}$. Because of the divergent behavior of $W^{\mathrm{c}} \sim 1/k^2$, the Brillouin-zone integration cannot be approximated by a finite summation as in Eq. \eqref{vB}. Therefore, we have replaced the ${\bf k}$ sum by an integral $\int_{\bf \Gamma}$, which could be understood as an integration over a finite region around ${\bf k}={\bf 0}$. In practice, we use instead an integration over the whole reciprocal space, not of $1/k^2$ (which would yield infinity), but of $\text{e}^{-\epsilon k^2}/k^2$ with a small positive coefficient $\epsilon$, and subtract a double-counting correction given by the sum over the ${\bf k}$-point set excluding the $\Gamma$ point. For details, see Ref. \citen{FLAPW} and in particular Eq. (34) therein. %The first term in \eqref{gamma} is uniform in ${\bf r}$ and ${\bf r}'$ to leading order, so that only the spherical average of $W_{11}^c({\bf k},\omega)$ survives integration. If we let $f(\omega)/(\sqrt{4\pi }k^{2})$ be the spherical average, the first term reduces to
%\begin{align}
%\frac{f(\omega)}{\sqrt{4\pi}\Omega}\int_{{\bf \Gamma}}
%d\mathbf{k}\frac{1}{k^{2}},
%\end{align}
%where the ${\bf \Gamma}$-point integral is defined as in Eq. (34) of Ref. \citen{FLAPW}.
\subsection{$W$ and $U$ in Time Domain: \\ Impulse and Step Response}
\label{time} 
%After computing $W$ in position representation we Fourier transform it to time domain to obtain $W({\bf r},{\bf r}';\tau)$, defined and discussed in Eq. \eqref{Wrrtime} in Section \ref{test}.
It is interesting to study the retarded interaction both related to the impulse response and the step response of a solid. The former is to linear order given by $W({\bf r},{\bf r}';\tau)$, and we show below that the latter is accessible from the same quantity. 

The interpretation of $W$ is provided in Sec. \ref{test}. Since it was obtained from linear response theory by replacing the external potential with the instantaneous Coulomb interaction, $v({\bf r}-{\bf r}')\delta(\tau)$, we here denote it by $W[\delta]$. $W[\delta]$ is connected to the impulse response of the system, and is obtained by a simple inverse Fourier transform of $W(\omega)$:
\begin{align}
W({\bf r},{\bf r}';\tau)[\delta] \equiv W({\bf r},{\bf r}';\tau) = \int \frac{d\omega}{2\pi} \text{e}^{-i\omega \tau } W({\bf r},{\bf r}';\omega).
\end{align}
$W(\omega)$ is here assumed to be retarded, but the $W(\omega)$ described in Sec. \ref{test} is time-ordered. For positive frequencies the time-ordered and retarded $W(\omega)$ are identical, but the former is an even function 
\begin{figure}[h]
\includegraphics[width=1.0\linewidth]{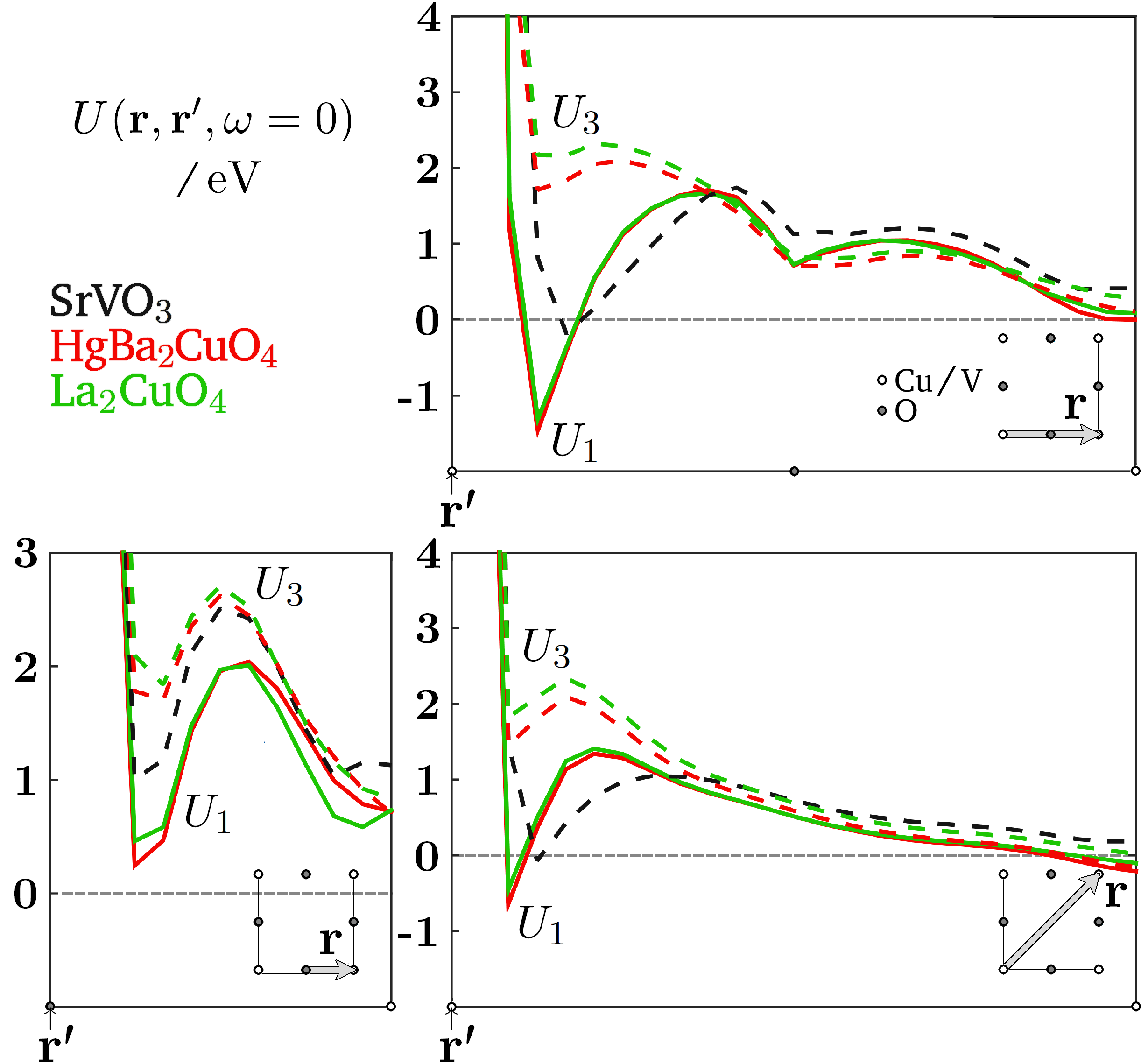}
\caption{Effective one- and three-band interaction, $U_1({\bf r},{\bf r}';\omega=0)$ (solid lines) and $U_3({\bf r},{\bf r}';\omega=0)$ (dashed lines), of the cuprates and $t_{2g}$ (three-band) interaction of SrVO$_3$ (black dashed lines) along different paths in the CuO$_2$ and VO$_2$ sheets respectively. These paths are indicated in each graph.}\label{U1D}
\end{figure}
of $\omega$ whereas the latter only has an even real part, but an odd imaginary part. By only calculating $W(\omega)$ for positive frequencies, the correct symmetries can easily be imposed. 

As is also clear from Sec. \ref{test}, if we instead introduce a point charge at ${\bf r}',t'$ kept frozen at later times, which means inserting $v({\bf r}-{\bf r}')\Theta(\tau)$ into Eq. \eqref{arbPert}, the resulting screened potential $W[\Theta]$ is given by 
\begin{align}
& W ({\bf r},{\bf r}';\tau)[\Theta]  = v({\bf r}-{\bf r}') \Theta(\tau)  \\
& + \int_0^\infty  d\tau_2 \int d{\bf r}_1 d{\bf r}_2 v({\bf r}_1 - {\bf r}_2) \chi({\bf r}_1,{\bf r}_2; \tau-\tau_2) v({\bf r}_2-{\bf r}') . \nonumber
\end{align}
Here, $\chi$ is the retarded response function, which is related to its time-ordered counterpart in the same way as described above for $W$. Since the retarded $\chi$ fulfills causality, the upper limit of integration can be changed to $\tau_2=\tau$, and from the variable substitution $\tau' = \tau -\tau_2$ we arrive at 
\begin{align}
& W ({\bf r},{\bf r}';\tau)[\Theta]  = v({\bf r}-{\bf r}') \Theta(\tau)  \\
& + \int_0^\tau  d\tau' \int d{\bf r}_1 d{\bf r}_2 v({\bf r}_1 - {\bf r}_2) \chi({\bf r}_1,{\bf r}_2; \tau') v({\bf r}_2-{\bf r}') \nonumber  \\
& = \int_{-\infty}^\tau d\tau' W({\bf r},{\bf r}';\tau')[\delta]. \nonumber 
\end{align}
This equation establishes a connection between the dynamically screened interaction between two electrons of the intrinsic system (impulse response) and the dynamically screened potential 
\begin{figure}[h!]
\includegraphics[width=1.0\linewidth]{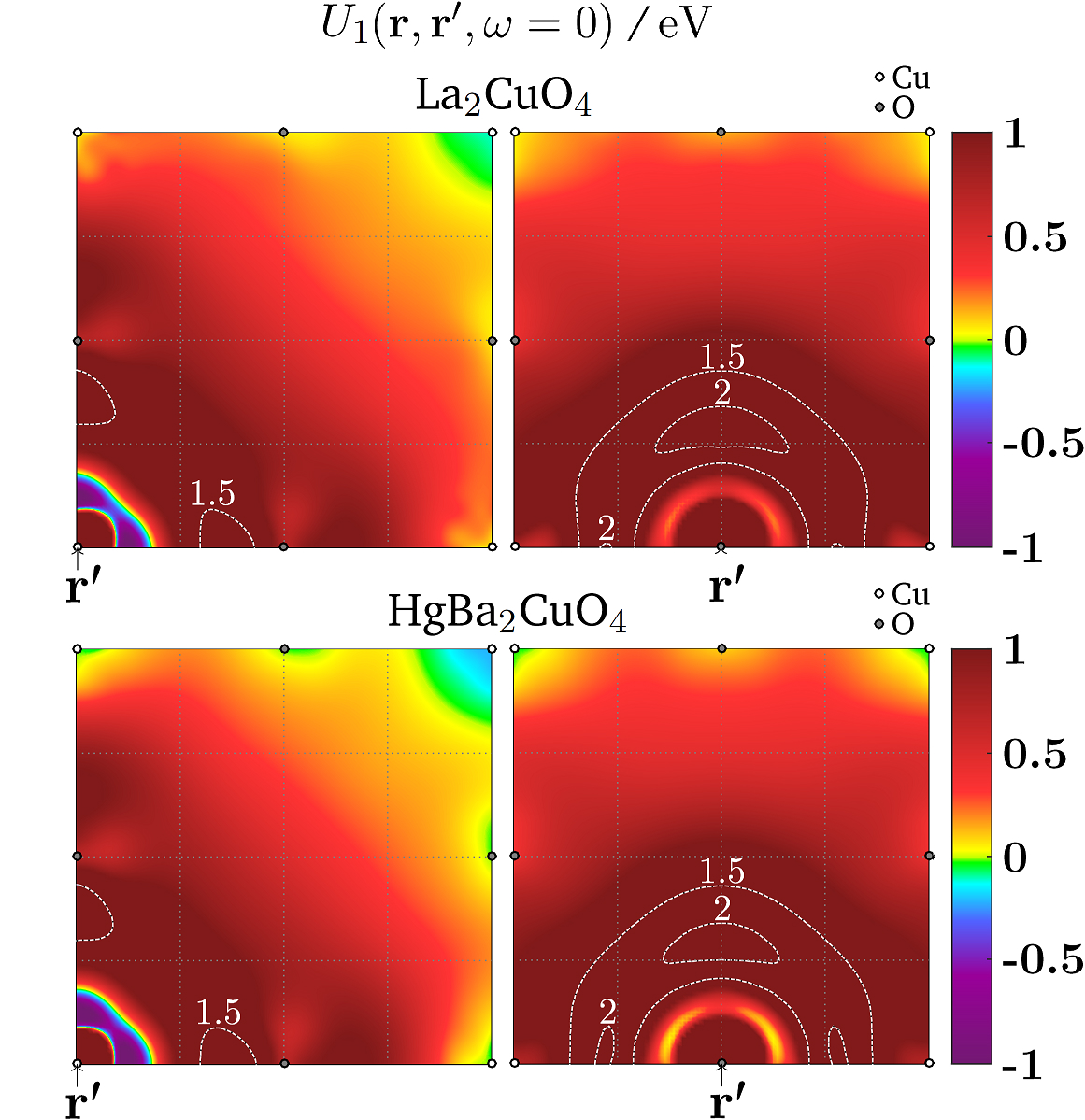}
\caption{Effective one-band interaction $U_1({\bf r},{\bf r}';\omega=0)$ of the cuprates in the CuO$_2$ sheet.}\label{U1}
\end{figure}  
from an impurity added to the system (step response). It has the following limits
\begin{align}
W({\bf r},{\bf r}';\tau)[\Theta]=\begin{cases}
v({\bf r}-{\bf r}') & ,~~~\tau \to 0^+ \\
W({\bf r},{\bf r}';\omega=0) & ,~~~\tau  \to \infty.
\end{cases} 
\end{align}

$W[\Theta]$ has dimension energy while $W[\delta]$ has dimension energy/time. 
\section{Results}\label{Sec:Results} 
We will now apply our method to compute the position representation of $W$ and $U$ in LCO, HBCO and non-superconducting SrVO$_{3}$. Computational details are provided in App. \ref{appendixB}.  We focus on the cases with $\mathbf{r}^{\prime}$ at the transition metal nucleus (Cu or V) as well as at the O nucleus, and with $\mathbf{r}$ and $\mathbf{r}'$ restricted to the same CuO$_2$ or VO$_2$ sheet. Furthermore, in all calculations, ${\bf r}$ and ${\bf r}'$ belong to the same unit cell.

\subsection{Static $U$ in Position Space}

%\begin{figure*}
%\includegraphics[width=.90\linewidth]{UDone.png}
%\caption{Effective one- and three-band interaction, $U_1({\bf r},{\bf r}';\omega=0)$ and $U_3({\bf r},{\bf r}';\omega=0)$, of the cuprates and t$_{2g}$ (also three-band) interaction of SrVO$_3$ in the CuO$_2$ and VO$_2$ sheets respectively.}\label{U}
%\end{figure*}

\begin{figure}[h]
\includegraphics[width=1.0\linewidth]{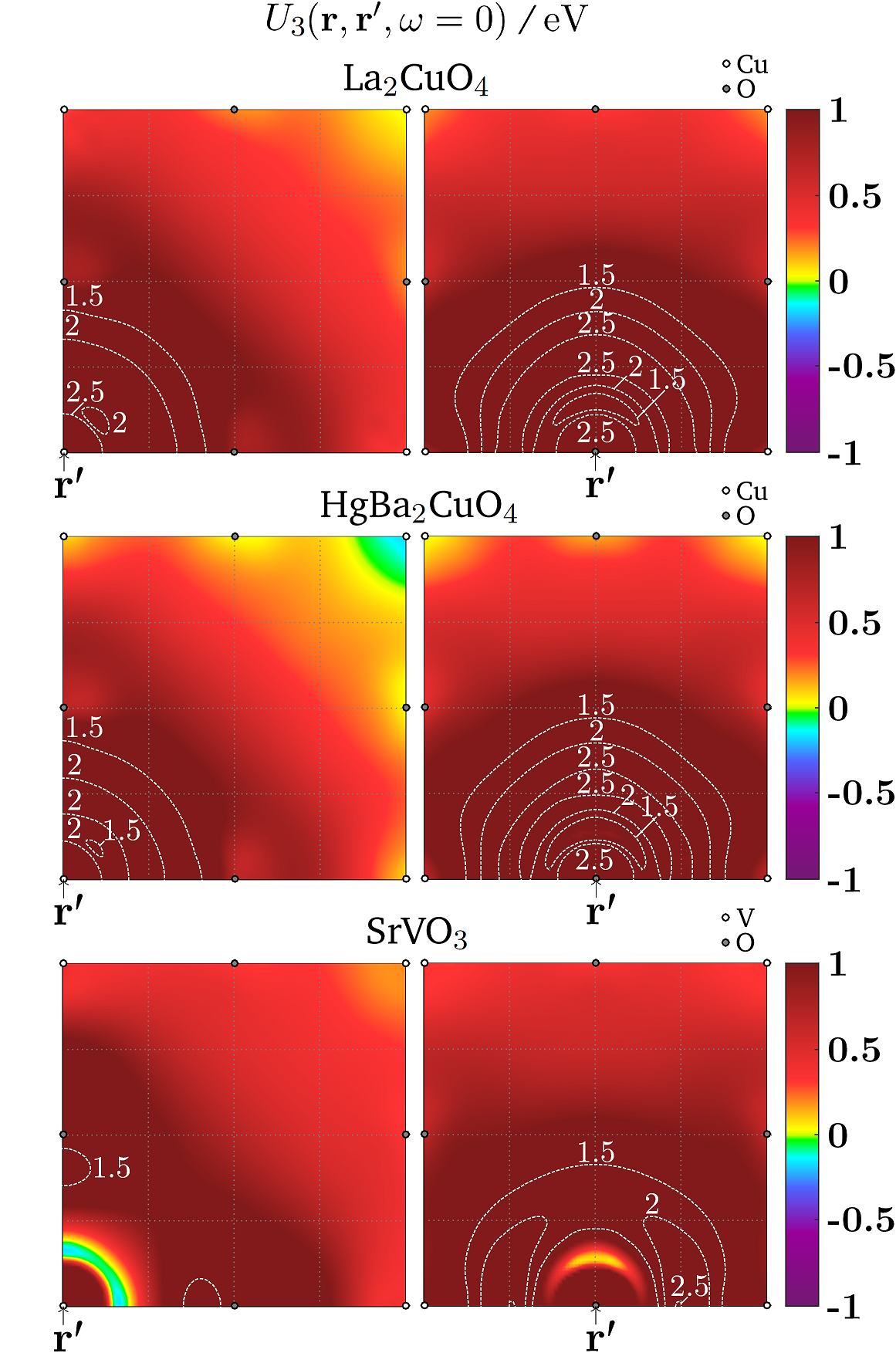}
\caption{Effective three-band interaction $U_3({\bf r},{\bf r}';\omega=0)$ of the cuprates and  SrVO$_3$ ($t_{2g}$) in the CuO$_2$ and VO$_2$ sheets respectively.}\label{U3}
\end{figure}

We start by considering the static effective interaction $U(\mathbf{r,r}^{\prime};\omega\!=\!0)$ (Fig. \ref{U1D} - \ref{U3}). We study the 1-band and 3-band models for the cuprates and compare the results with the non-superconducting perovskite SrVO$_3$ in the $t_{2g}$-model (see App. \ref{appendixB}). 

An interesting finding, with ${\bf r}'$ at the transition metal nucleus, is that the $t_{2g}$ interaction in SrVO$_{3}$ is essentially positive in the entire unit cell while in both cuprates there is a region close to the Cu site where $U_1$ ($U$ of the one-band model) is significantly negative. This region, as illustrated in Fig. \ref{U1}, has a shape which originates mainly from the 3$d_{x^2-y^2}$ orbital ($x^2-y^2$-derived) of the one-band subspace even though the intra-band screening from this orbital is excluded in the one-band model. Such a region does not appear in $U_3$ ($U$ of the three-band model) and thus originates from the hybridization between the Cu 3$d_{x^2-y^2}$ orbital and the O 2$p_x$ and 2$p_y$ orbitals. Since the $d$ orbitals are localized, this hybridization is expected to be strong only in their vicinity, which is consistent with the shape of the attractive region in $U_1$. However, while the $d_{x^2-y^2}$ orbital is antisymmetric with respect to a reflection of ${\bf r}=(x,y,0)$ across the line $x=y$, the attractive region is symmetric. This is physically clear, and can be understood from Eq. \eqref{vB}. If we let $\mathcal{R}$ be the reflection across $x=y$, we get
\begin{align}
W^{\mathrm{c}}(\mathcal{R} \mathbf{r},\mathbf{0};\omega)  &  = \sum_{ \mathbf{k} ,\alpha\beta}\mathcal{I}_{\alpha}^{\mathbf{k}%
}(\mathcal{R}{\bf r})\chi_{\alpha\beta}^{\text{RPA}}(\mathbf{k};\omega)\mathcal{I}%
_{\beta}^{\mathbf{k}\ast}(\mathbf{0}).
\end{align}
Since 
\begin{align}
    \mathcal{I}_\alpha^{\bf k}(\mathcal{R}{\bf r})& = \mathcal{I}_\alpha^{\mathcal{R}^{-1}{\bf k}}({\bf r})
\end{align}
and 
\begin{align}
 \chi_{\alpha \beta}^{\text{RPA}}({\bf k};\omega) &= \chi_{\alpha \beta}^{\text{RPA}}(\mathcal{R}^{-1} {\bf k};\omega)    
\end{align}
it follows that
\begin{align}
W^{\mathrm{c}}(\mathcal{R} \mathbf{r},\mathbf{0};\omega)  &  = W^{\mathrm{c}}({\bf r},{\bf 0};\omega).
\end{align}
A striking difference can be seen between the cuprates in the one-band model (Fig. \ref{U1}) and SrVO$_3$ in the $t_{2g}$ model (Fig. \ref{U3}). As already pointed out, in the cuprates, the region with strong one-band attraction coincides with the region with a large one-band density, which means that the electrons could feel the attraction. In SrVO$_3$, on the other hand, the region with the modest attraction in the minimal (three-band) $t_{2g}$ model, does not coincide with the region of the important in-plane $xy$ orbital of the model. This means that the electrons most likely experience repulsion. This finding is backed by earlier work\cite{firstHubb} on the screening channels that determine $U_3$ in SrVO$_3$, where if was found that O ${2p}-$V ${e_g}$ transitions constitute a stronger channel than O ${2p}-$V $t_{2g}$ transitions. 

It is also worth stressing, with ${\bf r}'$ at the Cu site, the negative $U_1$ at the next-nearest Cu site in both cuprates. The attraction is the strongest in HBCO, for which it survives in the three-band model. HBCO is also the only compound which displays attraction, though weak, at the neighboring Cu site (in the one-band model). The corresponding $t_{2g}$ interaction in SrVO$_3$ at the nearest or next-nearest neighbor V site is significantly positive. When $\mathbf{r}^{\prime}$ is moved to the O site the only identified attraction is very weak and found in the one-band model of HBCO at the next-nearest Cu site as can be seen in Fig. \ref{U1}. 

The matrix elements of the static $U_{1}$ in the maximally localized Wannier orbitals are positive for both cuprates\cite{laurentium,yang16Direct} but the observed region between the Cu and O sites with large negative $U_{1}$ opens up a possibility of having negative matrix elements in some other
orbitals, with a large weight in the attractive region. It is conceivable that such a basis could be used to describe possible Cooper pairs derived entirely from charge fluctuations. Such a basis cannot be found in non-superconducting SrVO$_{3}$ since the $U$ of the $t_{2g}$ model is almost entirely positive, at least in the first unit cell. 

In Sec. IV-C we analyze the screening channels associated with Cu 3$d_{x^2-y^2}-$3$d_{x^2-y^2}$ as well as O 2$p_{x,y}-$Cu 3$d_{x^2-y^2}$ transitions, but first we discuss the fully screened interaction $W$. 
\begin{figure}[h!]
\includegraphics[width=1.0\linewidth]{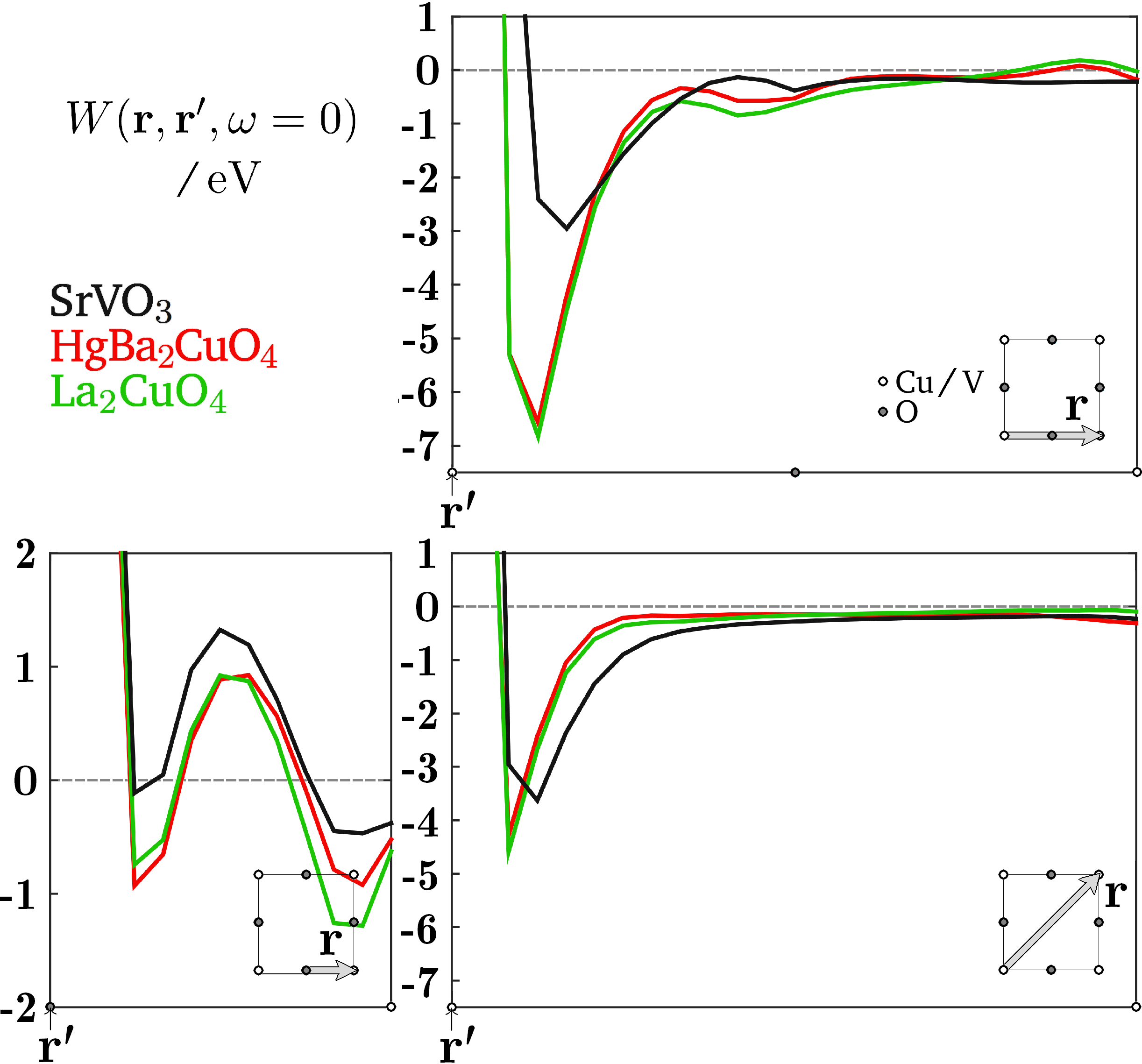}
\caption{$W({\bf r},{\bf r}';\omega=0)$ of the cuprates and SrVO$_3$ along different paths in the CuO$_2$ and VO$_2$ sheets respectively. These paths are indicated in each graph.}\label{W1D}
\end{figure}
\subsection{Static $W$ in Position Space}
$W$ contains all screening channels of the system, including, in the case of the cuprates, the spurious
metallic screening due to the pathological LDA band structures. The physical meaning of $W$ in this case is therefore not entirely clear. With this caveat in mind, it is nevertheless instructive to compute $W$ to understand the role of the screening within the antibonding band crossing the Fermi level, which may be thought of as modeling the screening of the
doped system. 

\begin{figure}[h!]
\includegraphics[width=1.0\linewidth]{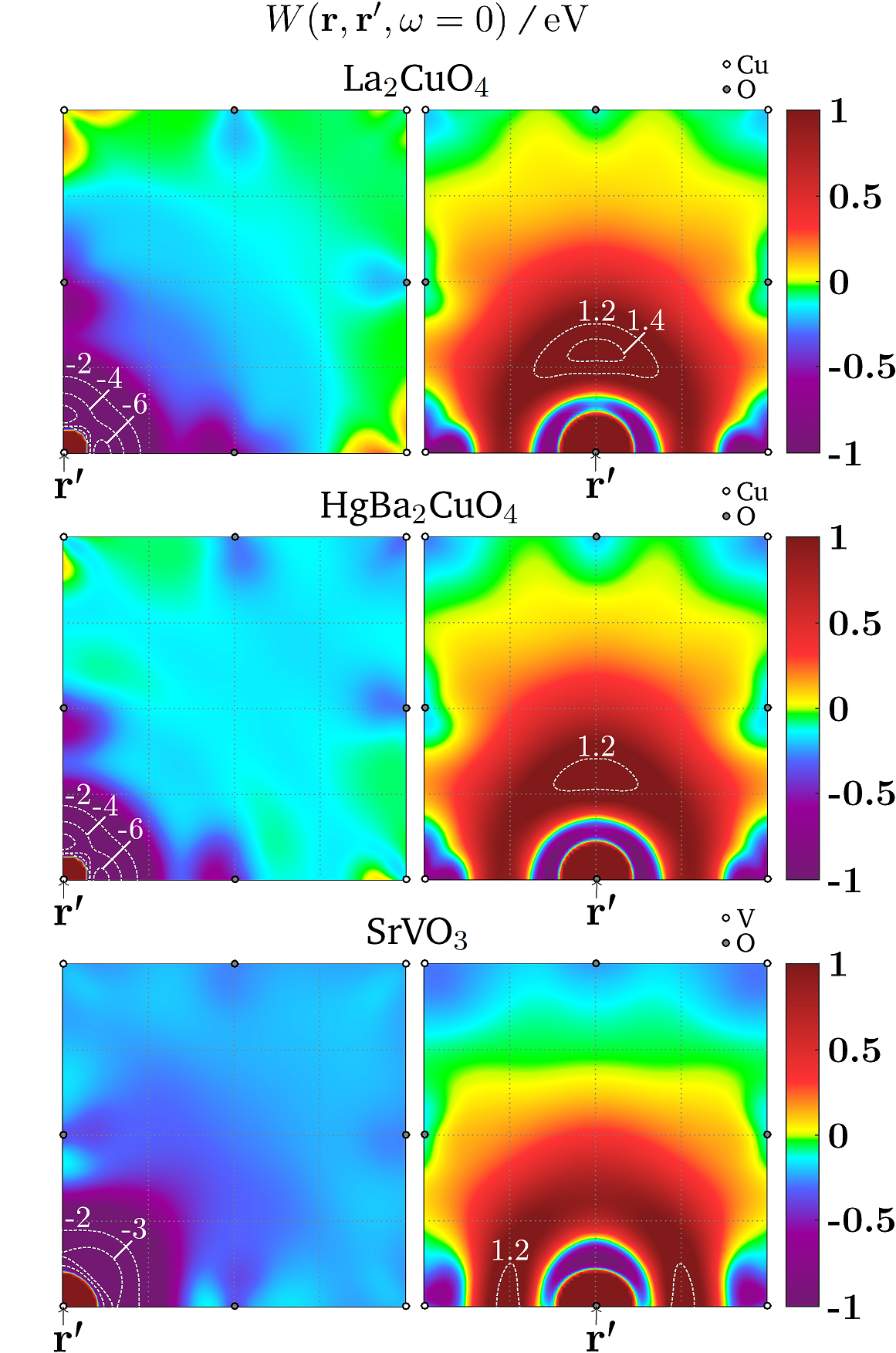}
\caption{$W({\bf r},{\bf r}';\omega=0)$ of the cuprates and SrVO$_3$ in the CuO$_2$ and VO$_2$ sheets respectively.}\label{W}
\end{figure}

In Fig. \ref{W1D} and \ref{W} we compare $W({\bf r},{\bf r}';\omega=0)$ in the CuO$_2$ sheets of the cuprates with that of the VO$_2$ sheet of SrVO$_3$. When choosing ${\bf r}'$ at the Cu or V site, large regions appear with negative $W$ in all of the compounds, but with a larger magnitude in the cuprates than in SrVO$_3$ (-6 versus -3 eV). This can be understood by observing that in the case of the cuprates, $W$ is obtained by screening $U_1$ with Cu 3$d_{x^2-y^2}-$3$d_{x^2-y^2}$ transitions, which have the same shape as $U_1$ itself. The screening in the $x^2-y^2$ channel is thereby enhanced. In SrVO$_3$, on the other hand, the screening in the $xy$ channel essentially only originates from within the $t_{2g}$ subspace, since there are no close-by orbitals outside the subspace to hybridize with.  

To investigate whether the $x^2-y^2$-derived shape of $W$, that can be seen in Fig. \ref{W}, is consistent with a superconducting gap with $x^2-y^2$ symmetry, we consider the superconducting DFT (SCDFT) gap equation. When excluding the effect of phonons, the SCDFT gap equation contains only the Kohn-Sham eigenenergies $\varepsilon_{n{\bf k}}$ and the static $W$ and reads\cite{scdft0,scdft}
\begin{align}
\Delta_{n}({\bf k})& = - \dfrac{1}{2}\sum_{n'{\bf k}'} W_{nn'}({\bf k}-{\bf k}';0) \dfrac{\tanh\bigg(\dfrac{\beta}{2} \mathcal{E}_{n'{\bf k}'} \bigg)}{\mathcal{E}_{n'{\bf k}'}} \Delta_{n'}({\bf k}') ,
\end{align}
where $\mathcal{E}_{n{\bf k}}  =((\varepsilon_{n{\bf k}}-\mu)^2+|\Delta_{n{\bf k}}|^2)^{1/2}$.
Furthermore, if $W$ or $\Delta$ have certain symmetries under a unitary transformation $S$ in position representation, this holds analogously in reciprocal space. Since we are only interested in the symmetry of the gap $\Delta$, we simplify the equation by linearizing it around $T=T_C$, where $\Delta_{n{\bf k}}$ is small. Since the ratio $\tanh(\frac{\beta}{2}\mathcal{E}_{n'{\bf k}'})/\mathcal{E}_{n'{\bf k}'}$ is a quickly decaying function we only keep the diagonal matrix element of $W$ from the band that crosses the Fermi level. We can then drop the band index completely and obtain 
\begin{figure}[h]
\includegraphics[width=1.0\linewidth]{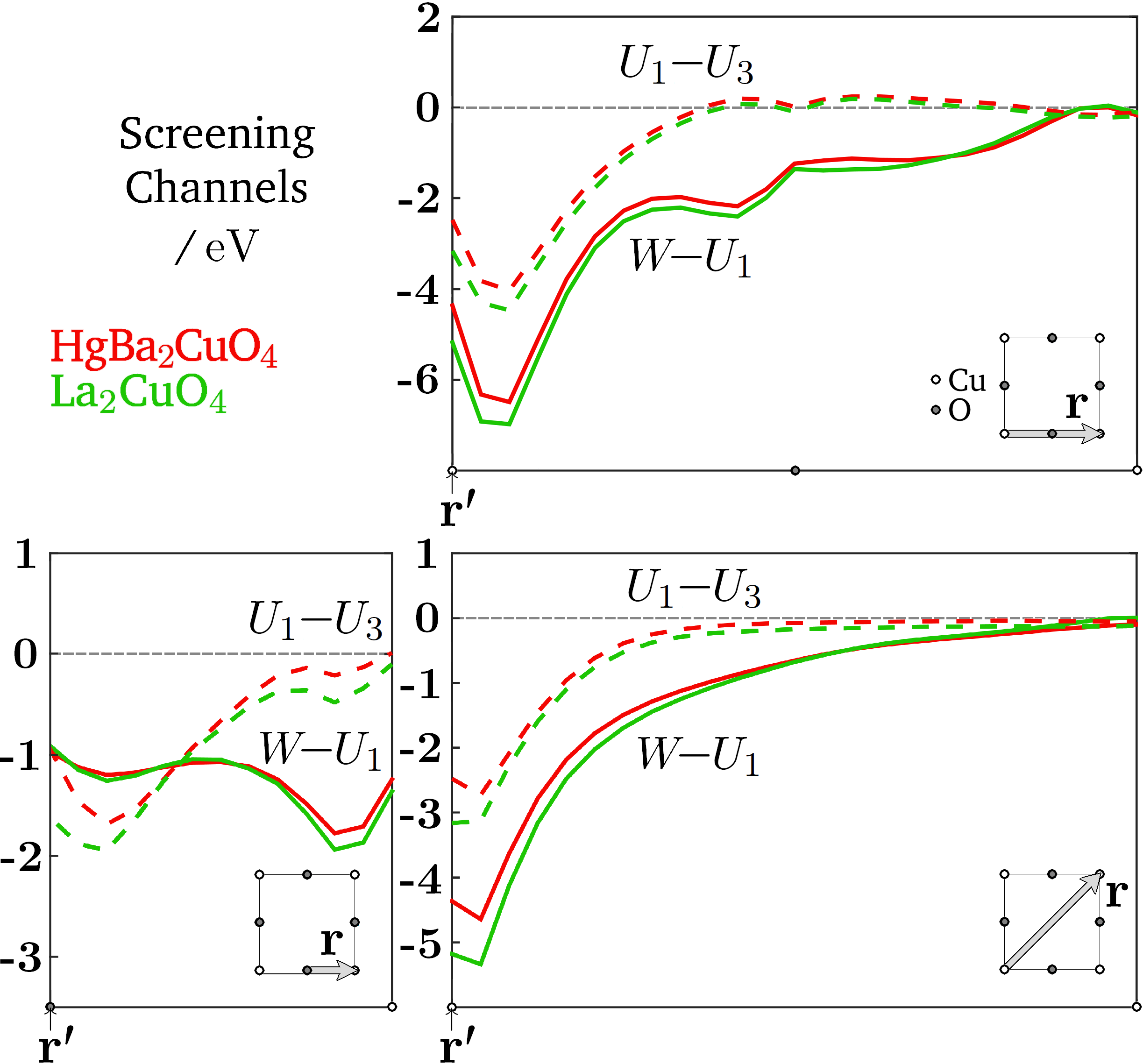}
\caption{$[W-U_1]({\bf r},{\bf r}';\omega=0)$ (solid lines) and $[U_1-U_3]({\bf r},{\bf r}';\omega=0)$ (dashed lines) in the cuprates along different paths in the CuO$_2$ sheet, which are indicated in each graph.}\label{Channel1D}
\end{figure}
\begin{align}
\Delta({\bf k}) & \approx - \frac{\beta}{4}\sum_{{\bf k}'} W({\bf k}-{\bf k}';0) \Delta({\bf k}').\label{SCeq}
\end{align}
The symmetry can now be deduced by considering a 3$\times$3 ${\bf k}$ mesh, corresponding to the first Brillouin zone, for which we make the posteriori ansatz
\begin{align}
W = \begin{pmatrix}
c & b & c \\
b & a & b \\
c & b & c
\end{pmatrix} ~~,~~~
\Delta = \begin{pmatrix}
0 & -\Delta & 0 \\
\Delta & \ 0 & \Delta \\
0 & -\Delta & 0
\end{pmatrix},
\end{align}
where the mid element corresponds to the $\Gamma$ point. 
By inserting this ansatz in \eqref{SCeq} and recalling that $\beta \approx 1/k_BT_C$ the relation
\begin{equation}
b \approx 2c-a-4k_BT_C
\label{simplerel}
\end{equation}
is obtained. Note that $T_C$ is the critical temperature obtained from $W$, which in general is smaller or equal to the true critical temperature, depending on what correlations are included (plasmons in this work). Since the $\Gamma$-point contribution, $a$, is in general large and positive for $W$ this relation means that a nonzero $\Delta$ is possible only for sufficiently negative $b$. This simplified condition should be applicable also to the spin-fluctuation mechanism. Equation \ref{simplerel} confirms that the calculated shape of $W$ is consistent with a superconducting gap of $x^2-y^2$ symmetry. A similar ansatz could be made in the $xy$ channel for SrVO$_3$, and it is plausible that the equivalent condition is not fulfilled since the strength of attraction in the $xy$ channel in SrVO$_3$ (Fig. \ref{W}) is only half that found in the $x^2-y^2$ channel in the cuprates. An unfulfilled condition implies that $\Delta$ is zero throughout, which obviously is true for non-superconducting SrVO$_3$. 
\begin{figure}[h]
\includegraphics[width=.95\linewidth]{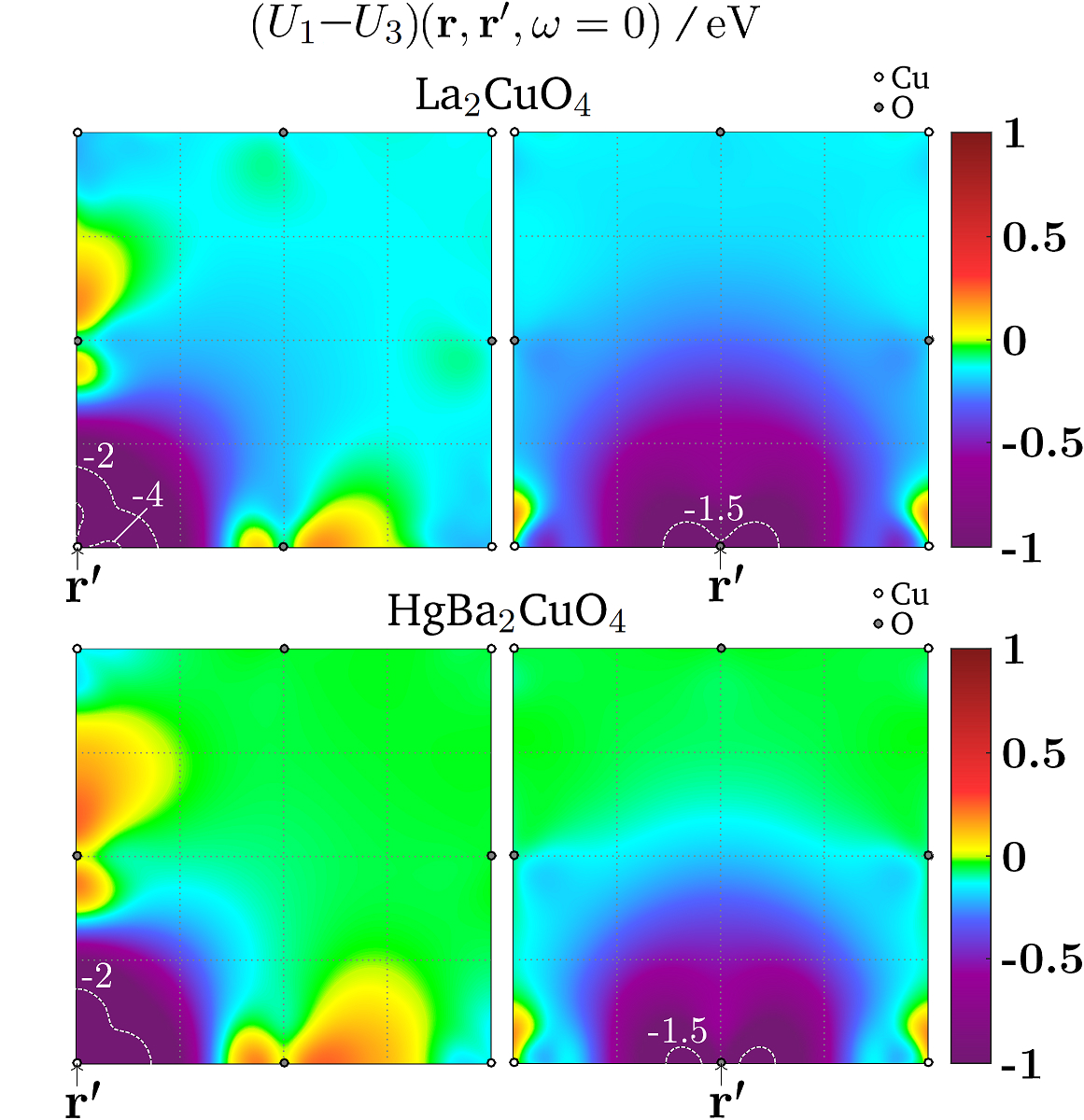}
\caption{$[U_1-U_3]({\bf r},{\bf r}';\omega=0)$ in the CuO$_2$ sheet of the cuprates.}\label{Channel2}
\end{figure}
\begin{figure}[h]
\includegraphics[width=.95\linewidth]{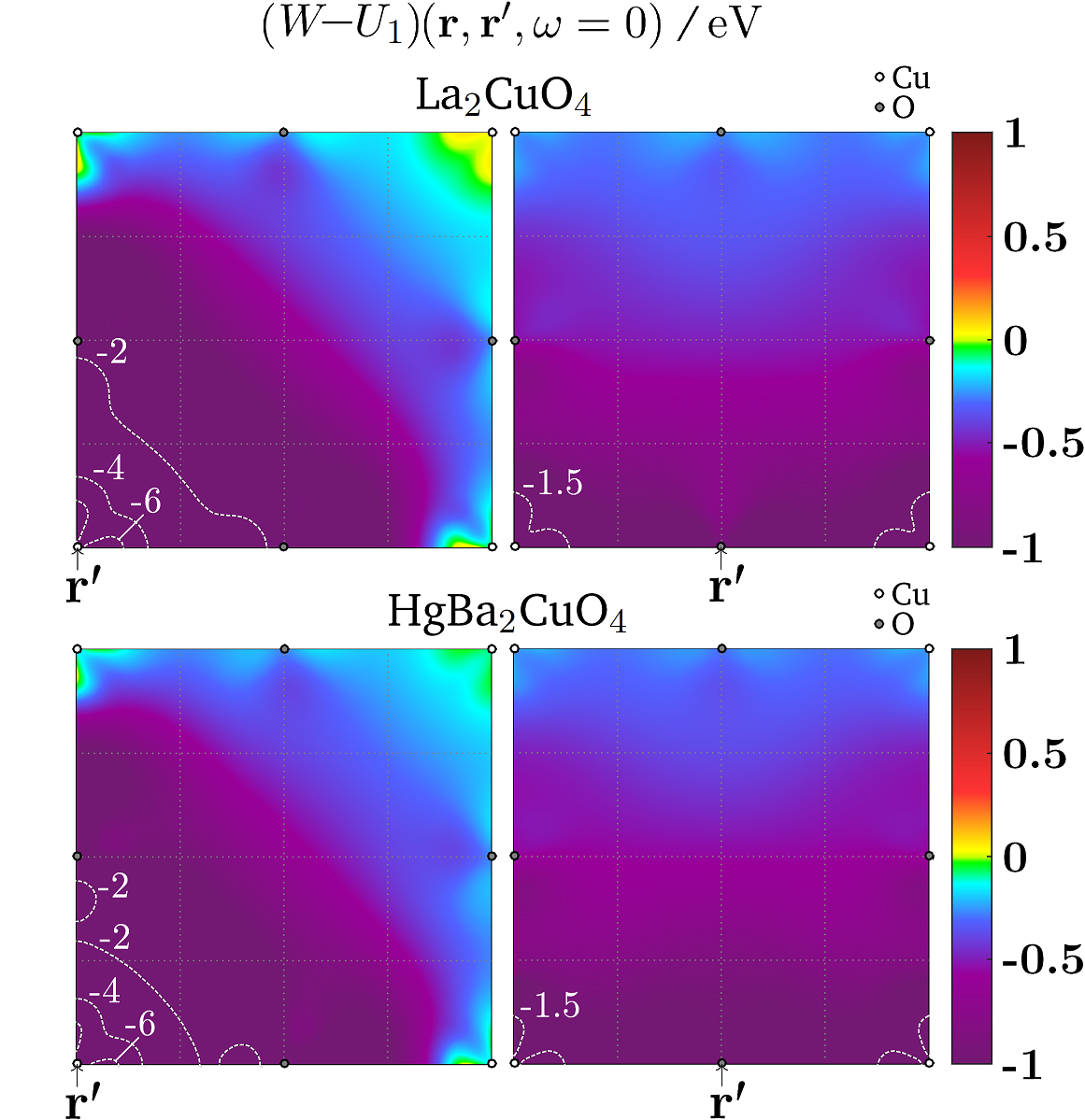}
\caption{$[W-U_1]({\bf r},{\bf r}';\omega=0)$ in the CuO$_2$ sheet of the cuprates.}\label{Channel1}
\end{figure}
\subsection{Screening Channels in Position Space}
%With $\chi^0_{\text{pd}}$ denoting polarization terms between occupied O 2p$_{x,y}$ and unoccupied Cu
%3d$_{x^{2}-y^{2}}$, we have $\chi_{1}^{\text{r}0}=\chi_{3}^{\text{r}0}+\chi^0_{\text{pd}}$,
%where $\chi_{i}^{\text{r}0}$ denotes $\chi^{\text{r}0}$ in the i-band model. Recalling
%that $U_{i}\!=\!v\!+\!v\chi_{i}^{\text{r}0}v\!+\!v\chi_{i}^{\text{r}0}v\chi_{i}%
%^{\text{r}0}v\!+\!\dots$ we see that
\begin{figure*}
\includegraphics[width=.62\linewidth]{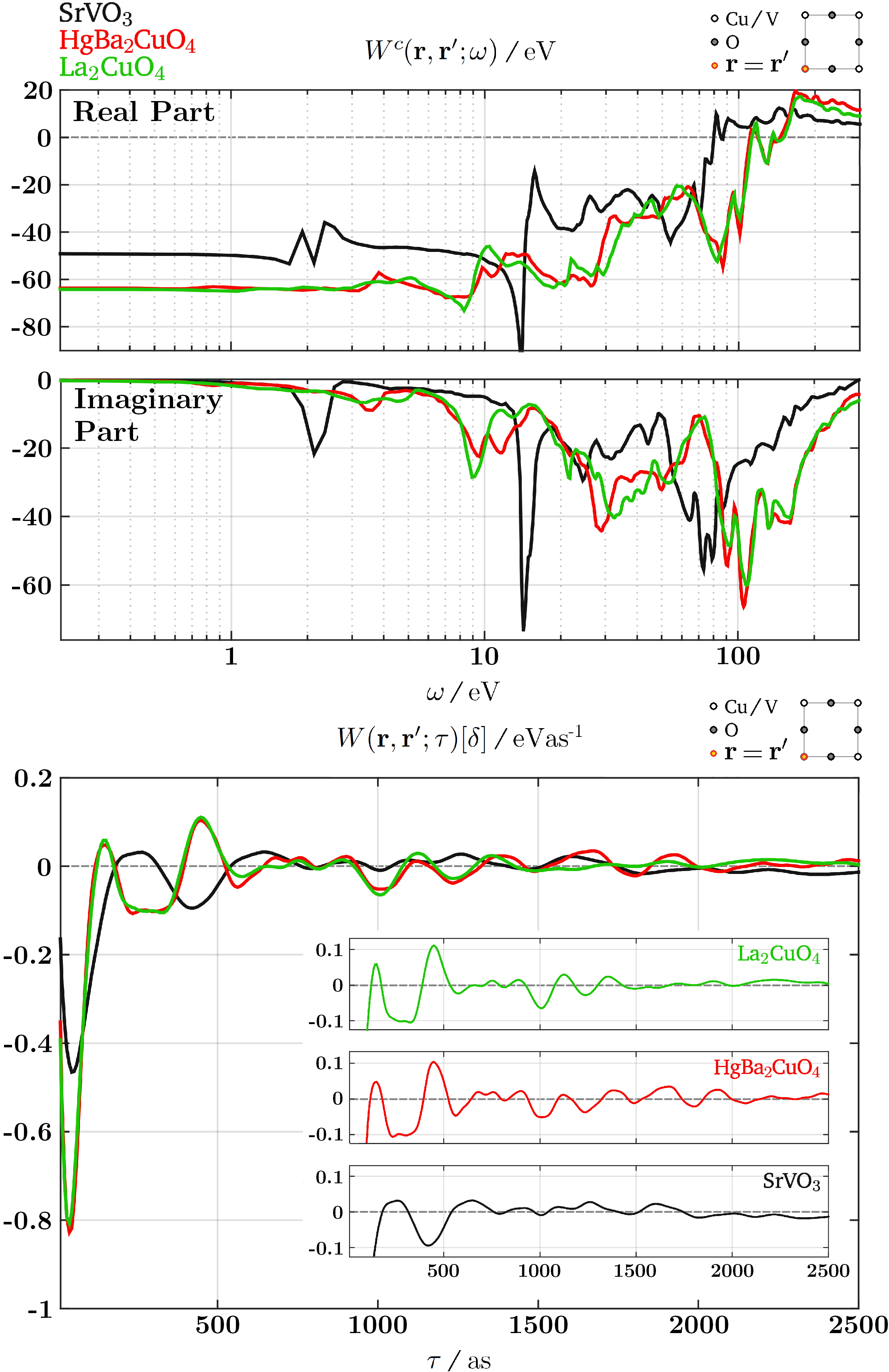}
\caption{$W^{\mathrm{c}}({\bf r},{\bf r}';\omega)$ and $W({\bf r},{\bf r}';\tau)[\delta]$ of the cuprates and SrVO$_3$ with ${\bf r}={\bf r}'$ at the Cu/V nucleus.}\label{Wtimedelta}
\end{figure*}
%\[
%U_{1}\!-\!U_{3}\!=\!v\chi^0_{\text{pd}}v\!+\!v\!\left[  \chi^0_{\text{pd}}v\chi^0_{\text{pd}%
%}\!+\! \chi^0_{\text{pd}}v\chi_{3}^{\text{r}0}+\chi_{3}^{\text{r}0}v\chi^0_{\text{pd}}\right]
%\!v\!+\!\dots
%\]

 Different polarization channels enter $\chi^\text{RPA}$ in a non-linear fashion. With the definition that the ''$pd$ screening'' comes from all terms in $\chi^\text{RPA}$ which contain O 2$p_{x,y}-$Cu 3$d_{x^2-y^2}$ transitions to linear order or higher, the resulting contribution to the effective interaction is exactly $U_1-U_3$ (Fig. \ref{Channel1D} and \ref{Channel2}). In the same manner, $W-U_1$ (Fig. \ref{Channel1D} and \ref{Channel1}) is the contribution from the ''$dd$ screening''. However, the Cu $d_{x^{2}-y^{2}}$ band in the 1- and 3-band models are not exactly identical. For this reason, in the computation of $U_1-U_3$, we calculate not only $U_3$ but also $U_{1}$ from the 3-band interpolation.

In agreement with earlier studies of LCO, \cite{laurentium} the $pd$ screening has most of its weight at the Cu site. It is clear from Fig. \ref{Channel1D} that the metallic $dd$ screening is stronger and has longer range than the $pd$ screening. The striking similarity between the results for LCO and HBCO indicate that the screening of the cuprates is generic, although the actual strength is material specific.
%We have also computed the fully screened interaction of doped iron pncitide
%FeSe as shown in Fig. The doped system is calculated within the virtual
%crystal approximation and the band structure as a function of doping is shown
%in Fig. Although the band structure exhibits a significant modification as a
%function of doping, surprisingly the screened interaction only shows a weak
%dependence on doping, which might indicate that electronic correlations do not
%play a direct role in superconductivity in these materials. On the other hand,
%the virtual crystal approximation employed in the present calculations may be
%too crude to capture subtle modifications in the electronic structure that can
%lead to substantial change in the screening process. A more reliable study
%should be carried out using a supercell approach, which is unfortunately
%computationally too demanding at present. As opposed to the virtual crystal
%approximation, the supercell approach would more faithfully reproduce the
%actual distribution of doping charge that can be very different from the
%periodic distribution assumed in the virtual crystal approximation. 
\begin{figure*}
\includegraphics[width=.62\linewidth]{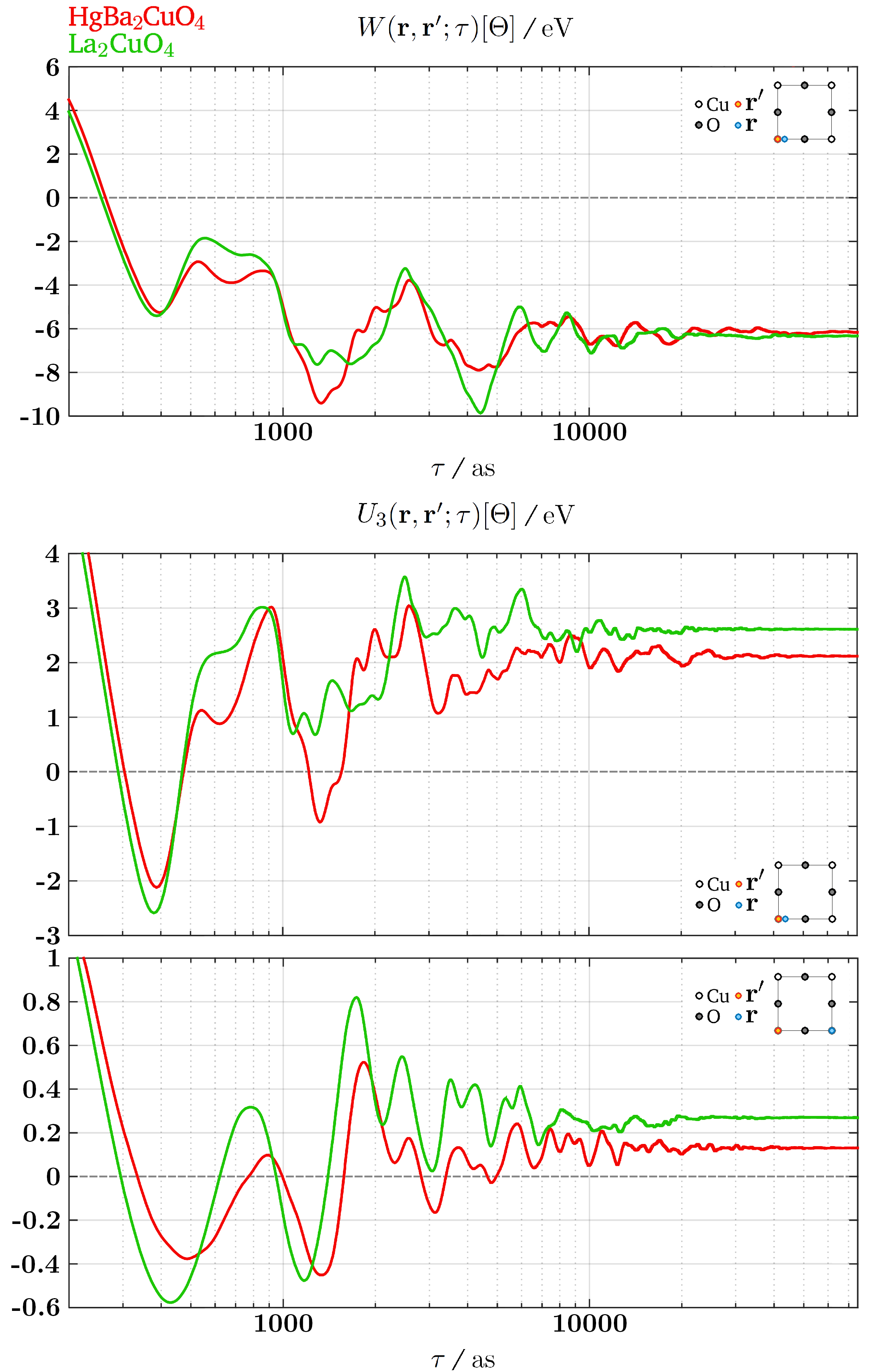}
\caption{$W({\bf r},{\bf r}';\tau)[\Theta]$ of the cuprates with ${\bf r}$ and ${\bf r}'$ at the same Cu nucleus as well as $U_3({\bf r},{\bf r}';\tau)[\Theta]$ with ${\bf r}$ and ${\bf r}'$ at the same Cu nucleus or at neighboring Cu nuclei.}\label{WtimeTheta}
\end{figure*}
\subsection{$W$ and $U$ in Time Domain}
The screened interaction $W({\bf r},{\bf r}';\tau)$ in time domain ($W[\delta]$ in Sec. \ref{time}) is presented in Fig. \ref{Wtimedelta} together with $W^{\mathrm{c}}({\bf r},{\bf r}';\omega)$ for LCO, HBCO and SrVO$_3$, with $\mathbf{r}$ and $\mathbf{r}^{\prime}$ at the same transition metal nucleus (Cu or V).

$W$ shares a common characteristic feature in time domain in all compounds. Shortly after the instantaneous bare interaction, there is a sudden surge of screening holes, which causes the large dip seen in $W$. $W$ then starts to oscillate, with a dominating characteristic frequency corresponding to the main collective charge excitation (plasmon) of
the system. This is superimposed by oscillations with different frequencies, corresponding to subplasmons of the system. Gradually, the oscillations decay and almost vanish after 2000 attoseconds. This can be understood
by considering the simple model ($\omega_n>0$)
\begin{align}
W^{\mathrm{c}}(\omega)  = -\frac{1}{\pi} & \sum_{n=1}^M  W_n \bigg[
\bigg(
\frac{\omega + \omega_n}{(\omega\!+\!\omega_n)^2\!+\!\delta^2} - \frac{\omega - \omega_n}{(\omega \!-\!\omega_n)^2 \!+\! \delta^2 }
\bigg) \nonumber \\
+& 
i\bigg(
\pi \delta(\omega \!+\! \omega_n)
+ \pi \delta(\omega \!-\! \omega_n)
\bigg) \text{sgn}(\omega)
\bigg],
\end{align}
where the imaginary part is
assumed to be a series of sharp $\delta$-functions, each representing a
subplasmon excitation with an appropriate weight $W_n>0$. Inverse Fourier transformation leads to
\begin{align}
W^{\mathrm{c}}(\tau) &= - \frac{2}{\pi} \sum_{n=1}^M W_n \sin(\omega_n \tau)\text{e}^{-\delta |\tau|} \Theta(\tau).
\end{align}
The behavior of $W^{\mathrm{c}}(\tau)$ for small $\tau$ is governed by the high-frequency features of $W^{\mathrm{c}}(\omega)$ and the dominating oscillation is determined by the bulk plasmon of the system. This explains the similar behavior for small $\tau$ in all the compounds in Fig. \ref{Wtimedelta} since the high-frequency electron gas-like bulk plasmon is usually present in real materials. Subplasmons of lower frequencies, on the other hand, are rather material specific and determine the behavior of
$W^{\mathrm{c}}(\tau)$ at large $\tau$. Indeed, in the time window between 1000 and
2000 attoseconds, $W^{\mathrm{c}}(\tau)$ still displays dramatic oscillations with strong attraction in both cuprates (mainly HBCO), but not in SrVO$_{3}$.

In Fig. \ref{WtimeTheta} we display the behavior of $W$ and $U_3$ in time-domain when an impurity is added to the system at $t=0$ and then left frozen at its position (see Sec. \ref{time}). As should be the case, the long-time limits equal the static ($\omega=0$) values of $W$ and $U_3$. $U_3$ is presented, but not $U_1$, because the static limit of the former is positive, whereas the static limit of $U_1$ is negative, just like that of $W$. The result for $U_3$ brings to light the presence of time intervals with a negative interaction, despite the static limit being positive. This shows the relevance of taking into account frequency dependence when utilizing $W$ or $U$ to model superconductivity.

\section{Summary and conclusions}
\label{Sec:Conclusions}
We have presented a method for computing the position representation of the effective electron-electron interaction $U$ in real materials and generalized the picture in time domain to include the study of static impurities. This basis-independent space-time approach is complementary to matrix element studies and allows for an unbiased perspective on the screening in real materials. This can be used to construct more suitable models of strongly correlated materials. 

As an illustration, we have applied the method within LDA cRPA to calculate the effective interactions in two well-known cuprate parent compounds, LCO and HBCO, as well as in the prototype of correlated metals, SrVO$_3$. We first studied the ${\bf r}$-dependence of $U({\bf r},{\bf r}';\omega=0)$, both with ${\bf r}'$ put at a transition metal nucleus (Cu or V) and at an in-plane O nucleus. In the $t_{2g}$ model of SrVO$_3$, with ${\bf r}'$ at the V nucleus, only a small region with weak attraction was found, which did not match the shape of the $xy$ low-energy orbital of the model. In the one-band model of the cuprates, on the other hand, a strong attractive interaction was found at the exact region of the low-energy 3$d_{x^2-y^2}$ orbital. Although this does not imply that charge fluctuations mediate Cooper pairing in the cuprates, they may assist other agents such as phonons and spin fluctuations in inducing pairing. 

The temporal interaction exhibited generic damped oscillations in all compounds. Its time integral was shown to be the potential caused by inserting an impurity at $\tau=0$, and the results for the three-band model illustrated the possibility of finite-time overscreening, with an attractive effective interaction, despite the static limit being repulsive. 

\begin{acknowledgments}
This work was supported by the Swedish Research Council.
\end{acknowledgments}

\begin{widetext}
\begin{figure*}[hb]
\includegraphics[width=.87\linewidth]{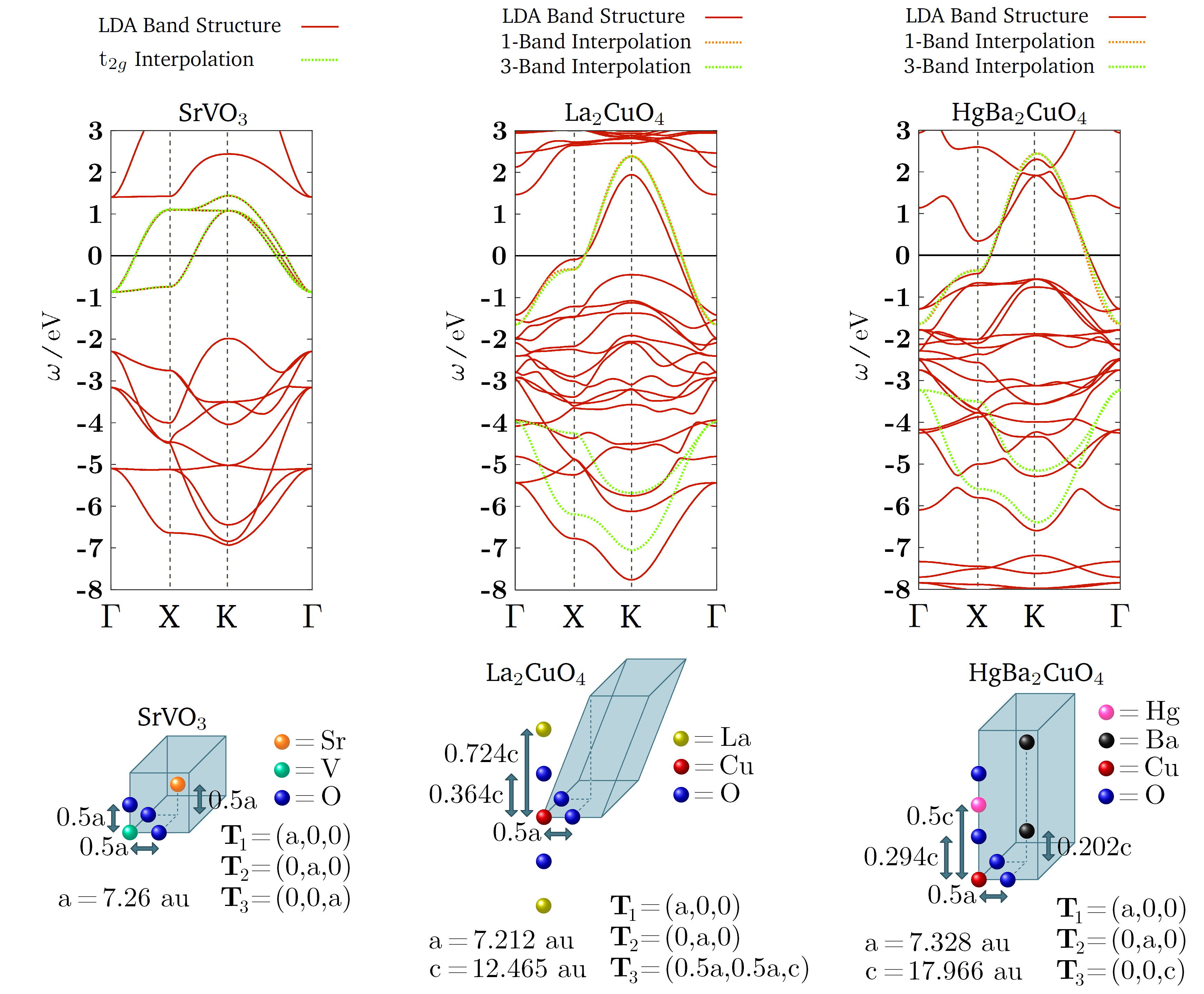}
\caption{LDA Band structures ($\mu$ at 0) and crystal structure data of SrVO$_3$, LCO and HBCO. $\Gamma=(0,0,0)$, $X=(\pi/a,0,0)$, $K=(\pi/a,\pi/a,0)$.}
\label{crystalband}
\end{figure*}
\end{widetext}
\appendix
\section{Computational Details}
\label{appendixB} 
We use the DFT code FLEUR\cite{{spexHP}} which utilizes the full-potential linearized augmented plane-wave (FLAPW) method to obtain all eigenfunctions $\phi_{n{\bf k}}$ and eigenvalues $\varepsilon_{n{\bf k}}$. All calculations are performed using the LDA. The band structures of HBCS, LCO and SrVO$_3$ are provided in Fig. \ref{crystalband} together with their crystal structures.\cite{latwocuofour,hg,srvothree} In LCO and HBCO we study $U$ in the well-established 1- and 3-band models, the former with a single Wannier function at Cu with $d_{x^{2}-y^{2}}$ symmetry and the latter also with two additional Wannier functions at the in-plane O atoms with $p_{x}$ and $p_{y}$ symmetry respectively. For comparison we also study $U$ in SrVO$_{3}$ in the $t_{2g}$ model, with three Wannier functions at V with $d_{xy}$, $d_{xz}$ and $d_{yz}$ symmetry. The Wannier interpolated band structures are provided together with the LDA band structures in Fig. \ref{crystalband}.

For the calculation of the RPA response matrix elements in the mixed product basis, $\chi_{\alpha \beta}^\text{RPA}({\bf k};\omega)$, we employ the SPEX code,\cite{spexHP} which uses the {\it ab initio} LDA eigensolution as the unperturbed mean-field reference system. The response matrix is then utilized to compute $W$ and $U$ in position representation in the way we have described in the present paper. Since the full frequency dependence is required for the calculation of the real-time dynamics, we have taken care to include all relevant screening processes, also virtual transitions from low-lying semicore states: Cu 3$p$ and V 3$p$. These states play an important role for large values of $\omega$, and it is indeed the 3$p$ local orbitals which are responsible for the large peak structures at around 100 eV in Fig. \ref{Wtimedelta}. In time domain, this only affects the first main interaction minimum. The interesting time interval around 1-2 fs is essentially unaffected. 

Surprisingly, the calculation turned out to be well converged with a sparse $4\times 4 \times 4$ ${\bf k}$-mesh. The effect of increasing the mesh-size to $8\times 8\times 8$ was minimal. All calculations are therefore performed using a ${\bf k}$ mesh of size 4$\times$4$\times$4. The CuO$_2$ sheets are, for simplicity, assumed to be perfectly two-dimensional without any buckling.

\end{document}